\documentclass[11pt,a4paper]{article}
\pdfoutput=1
\usepackage{jheppub}
\usepackage{graphicx}
\usepackage{amsmath, amssymb}
\usepackage{slashed}
\usepackage[table]{xcolor}
\usepackage[italicdiff]{physics}

\begin{document}

\preprint{KANAZAWA-24-06}

\title{Collective excitations
         in magnetic topological insulators 
         and axion dark matter search}

\author[a]{Koji Ishiwata}
\author[b]{Kentaro Nomura}

\affiliation[a]{Institute for Theoretical Physics, Kanazawa University,
  Kanazawa 920-1192, Japan}

\affiliation[b]{Department of Physics, Kyushu University, Fukuoka 819-0395, Japan}

\emailAdd{ishiwata@hep.s.kanazawa-u.ac.jp}
\emailAdd{nomura.kentaro@phys.kyushu-u.ac.jp}

\abstract{ We investigate collective excitations in magnetic
  topological insulators (TIs) and their impact on axion detection. In
  the three-dimensional TI model with the Hubbard term, the effective
  action of magnons and amplitude modes is formulated by dynamical
  susceptibility under the antiferromagnetic and ferromagnetic states.
  One of the amplitude modes is identified as ``axionic''
  quasi-particle and its effective coupling to the electromagnetic
  fields turns out to be roughly unchanged, or suppressed by up to two
  orders of magnitude, compared to the previous estimate, which may
  drastically change the sensitivity of the axion search using``axion''
  in magnetic TIs. }

\maketitle

\section{Introduction}
\label{sec:intro}

Axion and axion-like particles are candidates for dark matter of the
universe~\cite{Preskill:1982cy,Abbott:1982af,Dine:1982ah}. While many
observations have been dedicated to the search, it is yet to be found
so far. Recently, the dynamical `axion', as a quasi-particle in
condensed matter, has drawn attention in the context of the axion dark
matter search. The `axion' couples to the electromagnetic field and
Ref.\,\cite{Marsh:2018dlj} proposed an axion search in the mass range
of meV using the magnetic topological insulators (TIs).\footnote{See,
for examples,
Refs.\,\cite{Tokura2019,Bernevig2022,Liu2023,Sekine_2021}, for recent
reviews of magnetic TIs.  }

The mass scale is based on an estimation of the mass of the dynamical
`axion' in the TIs~\cite{Li:2009tca}. Bi$_2$Se$_3$, Bi$_2$Te$_3$, and
Sb$_2$Se$_3$ are the representative examples of three-dimensional (3D)
TIs.  Since the TIs do not have magnetic order, it is assumed in this
estimation that magnetic impurities, such as Fe or Cr, are doped to
Bi$_2$Se$_3$ so that the materials have ferromagnetism (FM) or
antiferromagnetism (AFM). Then Ref.\,\cite{Li:2009tca} proposes that
the magnetism is described by introducing the Hubbard interaction
term. Namely, the total model consists of the effective Hamiltonian
for the 3D TIs and the Hubbard term.  In the literature the `axion'
mass is estimated to be about 1~meV.  On the other hand, it was shown
in Ref.\,\cite{Ishiwata:2021qgd} that the typical mass of the `axion'
is of the order of eV, independent of the topology of insulators, in
the same model. The study also indicates that it can be suppressed
near the phase boundary of the AFM order. Similar results are reported
in a different model in which the spin interaction term between the
magnetic impurity and electron is
introduced~\cite{Ishiwata:2022mzq}. In the study the phase transitions
of the AFM and FM are shown and the `axion' mass is typically eV in
both phases.

Determination of the `axion' mass is crucial for the proposed axion
search. Furthermore, it is important to find a magnetic environment
that is suitable for the axion detection. In fact, the
first-principles calculation shows that Mn$_2$Bi$_2$Te$_5$, which is
recently synthesized experimentally~\cite{PhysRevB.104.054421}, has a
variety of magnetic states~\cite{Li:2020fvr}. In the literature the
topological phases are tuned by layer magnetization and the energy gap
is computed in each magnetic state.  Tuning of magnetism by an
electric field in magnetically doped TIs is discussed in
Ref.\,\cite{Wang_2015}.  On the other hand,
Ref.\,\cite{Lhachemi:2023eha} claims that magnetic field induces an
effective phonon charge, which plays a role of `axion.'  These studies
inspire us to search for other possible magnetic excitations that may
be utilized for the axion search.

In this paper, we investigate possible collective excitations in the
magnetic TI model and study their impact on the axion detection.  To
this end, we revisit the original model of the magnetic TIs, i.e., the
effective model for 3D TIs augmented by the Hubbard term.  Starting
from the partition function, the effective action for the excitations
is derived using the dynamical susceptibility. The emergence of the
AFM and FM orders is shown and we analyze the gap and dispersion
relation of eight magnetic excitations that consist of the amplitude
modes and magnons with various types.\footnote{The amplitude mode is
also called `Higgs amplitude mode' and various experimental efforts
have been made to detect it, for instance, in low-dimensional magnetic
materials~\cite{PhysRevLett.85.832,PhysRevLett.89.197205,
  PhysRevLett.93.257201,PhysRevLett.100.205701,Jain_2017,Hong_2017} or
frustrated magnetic
materials~\cite{PhysRevB.97.140405,osti_1571849}. For the theory of
the spin wave, see, for example,
Refs.\,\cite{PhysRev.81.869,PhysRev.85.1003,PhysRev.87.60,PhysRevB.64.184423,PhysRevB.93.094438}.}
We identify one of the amplitude modes as `axion' and one of the
magnons also can play the role of `axion' in a specific case. The
validity of the effective description of the excitations is clarified
and some excitations are found to be unstable.

The paper is organized as follows. The effective action is derived
from the model in Sec.\,\ref{sec:model}. From the effective action the
effective potential for the order parameters of the AFM and FM is
calculated in Sec.\,\ref{sec:Veff}, and the features of the magnetic
orders are discussed. In Sec.\,\ref{sec:excitations} the effective
actions of the eight types of collective excitations are derived and
the dispersion relation and the stability of the excitations are
described. Here we identify the `axion' quasi-particle and investigate
its properties. Based on the results, we discuss a direction to
realistic axion search in Sec.\,\ref{sec:axion_search}.
Sec.\,\ref{sec:conclusion} is devoted to conclusion.

In our study, we adopt natural units where $\hbar=c=k_B=1$.

\section{The model}
\label{sec:model}

To discuss possible magnetic orders, we consider an effective
Hamiltonian for 3D TIs~\cite{Zhang:2009zzf,Li:2009tca,Liu_2010} that
is augmented by the Hubbard term. The Hamiltonian of the model is
given by
\begin{align}
  H=H^{\rm TI}+H_U\,,
\end{align}
where
\begin{align}
  H^{\rm TI} &= \sum_{i} c^\dagger_i{\cal H}^{\rm TI} c_i\,,
  \\
  H_U &= U\sum_{I= A,B}\sum_{i} n_{Ii\uparrow}n_{Ii\downarrow}\,.
\end{align}
Here $U$ is a positive constant, $I=A,B$ and $i$ are the indices of
the sublattice and site, respectively.
$n_{Ii\sigma}=c^\dagger_{Ii\sigma}c_{Ii\sigma}$ is the number density
of the electrons at the site with spin $\sigma=\uparrow,\downarrow$
and
$c_i=(c_{Ai\uparrow},c_{Ai\downarrow},c_{Bi\uparrow},c_{Bi\downarrow})^T$.
${\cal H}^{\rm TI}$ will be given later.  In our study, we consider
$z$ axis as the easy axis.  By the Stratonovich-Hubbard
transformation, the Hubbard term is written by introducing two fields
$\phi_{Ii}$:
\begin{align}
  H_U \to
  &\frac{U}{4}\sum_{I,i}\left[
    \phi^2_{Ii}+2\phi_{Ii}(n_{Ii\uparrow}-n_{Ii\downarrow})
    \right]
  \label{eq:H_U_1} \\
  =& \frac{U}{2}\sum_{i}(\phi^2_{fi}+\phi^2_{ai})+
  \frac{U}{2}\sum_{i}c^\dagger_i\left[
  \phi_{ai}\Gamma^5+\phi_{fi}\Gamma^{12}
  \right]c_i\,,
  \label{eq:H_U_1a}
\end{align}
where $\Gamma^{12}$ and $\Gamma^5$ are 4 by 4 matrices defined as
\begin{align}
  \Gamma^{5}=\mqty(\sigma^3 &0 \\ 0 &-\sigma^3)\,,~~~~
  \Gamma^{12}=\mqty(\sigma^3 &0 \\ 0 &\sigma^3)\,.
\end{align}
Here $\vec{\sigma}=(\sigma^1,\sigma^2,\sigma^3)$ are the Pauli
matrices.  Appendix~\ref{app:Gammas} gives the list of the Gamma
matrices that are used in this study.  $\phi_{Ii}$ is physically the
spin moment in $z$ direction at site $i$ of sublattice $I$ and we have
introduced
\begin{align}
  \phi_{ai}\equiv (\phi_{Ai}-\phi_{Bi})/2\,, \\
  \phi_{fi}\equiv (\phi_{Ai}+\phi_{Bi})/2\,.
\end{align}
When $\phi_{ai}$ and $\phi_{fi}$ are constants, then their constant
values correspond to the order parameters of the AFM and FM,
respectively. The term proportional to $\Gamma^{12}$ is considered in
Refs.\,\cite{Rosenberg_2012, Kurebayashi_2014, Wang:2015hhf} in
different models. Then (Euclidean) action is given by
\begin{align}
  S=\int^\beta_0d\tau \sum_i\left[
    c^\dagger_i(\partial_\tau +{\cal H})c_i
    +\frac{U}{2}(\phi^2_{ai}+\phi^2_{fi})
    \right]
\end{align}
where $\tau=it$ is the imaginary time and
\begin{align}
  {\cal H}={\cal H}^{\rm
    TI}+(U/2)(\phi_{ai}\Gamma^5+\phi_{fi}\Gamma^{12})\,.
\end{align}
$\beta=T^{-1}$ is the inverse temperature. The effective action
$S_{\rm eff}$ is given by integrating out the electron field $c_i$ in
the partition function $Z$, i.e.,
\begin{align}
  Z&=\int {\cal D}c{\cal D}c^\dagger{\cal D}\phi_a{\cal D}\phi_f
  e^{-S}
  \nonumber \\
  &= \int {\cal D}\phi_a{\cal D}\phi_f
  e^{-S_{\rm eff}}\,,
  \label{eq:Z}
\end{align}
which leads to
\begin{align}
  S_{\rm eff}=\int^\beta_0d\tau \frac{U}{2}\sum_{i}(\phi^2_{ai}+\phi^2_{fi})
  -\ln \det (\partial_\tau +{\cal H})\,.
  \label{eq:Seff_orig}
\end{align}
We will use $Z$ and $S_{\rm eff}$ to derive the effective potential
for the order parameters in the next section.

\section{Magnetic orders}
\label{sec:Veff}

In this section we find possible global orders of the magnetism and
identify the ground state.  For this purpose, we ignore the local
quantum fluctuations and compute the effective potential following the
technique of Ref.\,\cite{Ishiwata:2022mzq}.

\subsection{The effective potential}

The effective potential is derived from the grand potential.  The
grand potential $\Omega$ is defined by the partition function
$Z$:\footnote{Though we give the formulae with finite temperature, we
will focus on the zero temperature limit in the later numerical
study.}
\begin{align}
  \Omega=-\beta^{-1}\ln Z\,.
\end{align}
The effective potential per site is then obtained by taking
$\phi_{ai}$ and $\phi_{fi}$ as constants, i.e., $\phi_{ai}=\phi_{a}$
and $\phi_{fi}=\phi_{f}$, 
\begin{align}
  V_{\rm eff}&\equiv -\beta^{-1}
  \ln e^{-S_{\rm eff}}/N|_{(\phi_{ai},\phi_{fi})=(\phi_a,\phi_f)}
  \nonumber \\
  &=\frac{U}{2}(\phi^2_{f}+\phi^2_{a})
  -\frac{1}{\beta N}\sum_{h,\vb*{k}}
  \ln (1+e^{-\beta (E_{h\vb*{k}}-\mu)})\,,
\end{align}
where $N$ is the number of the cite, $\mu$ is the chemical potential,
$h$ is the band index, and $\vb*{k}$ is the wavenumber. $E_{h\vb*{k}}$
will be given soon.  In the derivation, we have adopted the Matsubara
formalism, which gives
\begin{align}
  \ln \det (\partial_\tau +{\cal H}) =
  \sum_{n}\sum_{h,\vb*{k}}\ln (-i\omega_n +E_{h\vb*{k}}-\mu)\,,
\end{align}
where $\omega_n=(2n+1)\pi/\beta$ is the Matsubara frequency for
fermion.  $E_{h \vb*{k}}$ are the eigenvalues of ${\cal H}_{\vb*{k}}$,
which is ${\cal H}$ in the wavenumber space:
\begin{align}
  {\cal H}_{\vb{k}}={\cal H}^{\rm TI}_{\vb{k}}
  +d_5 \Gamma^5+d_{12}\Gamma^{12}\,,
\end{align}
where
\begin{align}
  {\cal H}^{\rm TI}_{\vb{k}}&=
  \epsilon_0{\bf 1}
  +\sum_{a=1}^4d_a\Gamma^a\,, \\
  (d_5,d_{12}) &= (U/2)(\phi_{a},\phi_f)\,.
\end{align}
$\Gamma^a$ ($a=1$\,-\,4) are the Gamma matrices given by
\begin{align}
  &\Gamma^{1}=\mqty(\sigma^1 & 0 \\ 0 & -\sigma^1)\,,~
  \Gamma^{2}=\mqty(\sigma^2 & 0 \\ 0 & -\sigma^2)\,,~
  \Gamma^{3}=\mqty(0 & -i\bf{1} \\ i\bf{1} &0)\,,~
  \Gamma^{4}=\mqty(0 & -\bf{1} \\ -\bf{1} &0)\,,
\end{align}
and we take $\phi_{a},\phi_{f}>0$ without loss of generality.
$\epsilon_0$ is a constant and $d^a$ is parametrized
as~\cite{Zhang:2009zzf,Li:2009tca,Liu_2010}
\begin{align}
  &(d_1,d_2,d_3,d_4) =
  \nonumber \\
  &~~~~~~~(A_2 \sin k_x a_x,\, A_2 \sin k_y a_y,
  \,A_1 \sin k_z a_z,\,{\cal M})\,,
  \label{eq:d_a}
\end{align}
where ${\cal M}=M_0-2B_1-4B_2+2B_1 \cos k_z a_z+2B_2(\cos k_x a_x+
\cos k_y a_y)$.  In the following study, we consider a cubic lattice
and use dimensionless wavenumber for simplicity, i.e.,
$a_x=a_y=a_z=1$. The first and second terms of ${\cal H}_{\vb*{k}}$
have the time-reversal invariance, which is one of the features of the
TIs, and it describes Bi$_2$Se$_{3}$ family of materials. The
eigenvalues are obtained as
$(E_{1\vb*{k}},E_{2\vb*{k}},E_{3\vb*{k}},E_{4\vb*{k}})
=(\epsilon_0-\epsilon_{1\vb*{k}},\epsilon_0-\epsilon_{2\vb*{k}},
\epsilon_0+\epsilon_{1\vb*{k}},\epsilon_0+\epsilon_{2\vb*{k}})$, where
\begin{align}
  \epsilon_{1\vb*{k}}&=
  \sqrt{d_0^2-d_s^2+[d_{12}+(d_s^2+d_5^2)^{1/2}]^2}\,,
  \\
  \epsilon_{2\vb*{k}}&=
  \sqrt{d_0^2-d_s^2+[d_{12}-(d_s^2+d_5^2)^{1/2}]^2}\,,
\end{align}
Here $d_0\equiv (\sum_{a=1}^4 d_a^2)^{1/2}$ and $d_s\equiv
(d_3^2+d_4^2)^{1/2}$.  Since we focus on the insulated states, we
consider that the electrons are half-filled. Namely, we take
$\epsilon_0-\mu\simeq 0$ and two lower bands $E_{1\vb*{k}}$ and
$E_{2\vb*{k}}$ are filled.

For later analysis, we derive the stationary conditions for the
possible magnetic orders. For the AFM, it is given by $\pdv*{V_{\rm
    eff}}{\phi_a}|_{(\phi_a,\phi_f)=(\phi_{a0},0)}=0$, and the result
is
\begin{align}
  1+\frac{U}{2N}\sum_{\vb*{k}} \{-n_F(-d)+n_F(d)\}\frac{1}{d}=0\,,
  \label{eq:phi_a0}
\end{align}
where $\phi_{a0}$ is a nonzero stationary value, $d\equiv
(\sum_{a=1}^5 d_a^2)^{1/2}$ and $n_F$ is the Fermi-Dirac distribution
function.  Similarly, $\pdv*{V_{\rm
    eff}}{\phi_f}|_{(\phi_a,\phi_f)=(0,\phi_{f0})}=0$ gives a nonzero
$\phi_{f0}$ that satisfies 
\begin{align}
  1+\frac{1}{2N\phi_{f0}}\sum_{\vb*{k}}
  \Bigr[\{-n_F(-\epsilon_1)+n_F(\epsilon_1)\}
    \frac{d_{12}+d_s}{\epsilon_1}
        +\{-n_F(-\epsilon_2)+n_F(\epsilon_2)\}\frac{d_{12}-d_s}{\epsilon_2}
    \Bigl]=0\,.
    \label{eq:phi_f0}
\end{align}
For both the AFM and FM cases, it is easy to check that
$\phi_{a0},\phi_{f0} \to 1$ as $U\to \infty$, which corresponds to a
case where the magnetism is saturated.

Before computing the effective potential, we discuss conditions for
the AFM and FM orders. To this end we take $T\to 0$ limit, which leads
to
\begin{align}
  V_{\rm eff}|_{T=0}=\frac{U}{2}(\phi^2_{f}+\phi^2_{a})-
  \frac{1}{N}\sum_{\vb*{k}}(\epsilon_{1\vb*{k}}+\epsilon_{2\vb*{k}})\,.
\end{align}
Regarding the AFM, we expand the effective potential up to
$\order{\phi_a^2}$ while taking $\phi_f=0$ to obtain
\begin{align}
  V_{\rm eff}|_{T=0,\phi_f=0}=\frac{U}{2}\phi_a^2
  \left[1-\frac{U}{2N}\sum_{\vb*{k}}\frac{1}{d_0}\right]
  +\order{\phi_a^4}+{\rm const.}
\end{align}
Therefore we expect to have the AFM order when
\begin{align}
  1-\frac{U}{2N}\sum_{\vb*{k}}\frac{1}{d_0} <0\,.
  \label{eq:PT_AFM}
\end{align}
Similarly, one may get a condition to have the FM order as
\begin{align}
  1-\frac{U}{2N}\sum_{\vb*{k}}\frac{d_1^2+d_2^2}{d_0^3} <0\,.
  \label{eq:PT_FM}
\end{align}
However, it is not an exact condition for the FM order; the FM order
may arise even when \eqref{eq:PT_FM} is not satisfied. This can be
understood by expanding the effective potential at $\order{\phi_f^4}$:
\begin{align}
   V_{\rm eff}|_{T=0,\phi_a=0}=\frac{U}{2}\phi_f^2
   \left[1-\frac{U}{2N}\sum_{\vb*{k}}\frac{d_1^2+d_2^2}{d_0^3}\right]
   + \frac{U^4\phi_f^4}{64}\sum_{\vb*{k}}
   \frac{d_0^4-6d_0^2d_s^2+5d_s^2}{d_0^7}
   +\order{\phi_f^6}\,.
   \label{eq:Veff_f}
\end{align}
The coefficient of a term proportional to $\phi_f^4$ can be negative,
which may lead to the FM order. We will confirm this numerically
later.  Even though it is not an exact condition, \eqref{eq:PT_FM} can
be used to understand the qualitative behavior of the emergence of the
magnetic order. I.e., we expect that the FM order will appear for
relatively larger $d_1$ and $d_2$ than $d_3$ and $d_4$, in other
words, larger $A_2$ compared to $A_1$, $B_1$, $B_2$, and $M_0$.

On the other hand, if $d_1=d_2=0$ is exactly satisfied, the effective
potential becomes
\begin{align}
  V_{\rm eff}|_{T=0,\phi_a=0}=\frac{U}{2}\phi_f(\phi_f-2)\,.
  \label{eq:Veff_F_A2=0}
\end{align}
Thus, the FM order appears without other conditions, and the stationary
point is given by $\phi_{f0}=1$. We will discuss this specific case
later.

Both conditions \eqref{eq:PT_AFM} and \eqref{eq:PT_FM} are satisfied
in the limit $U \to \infty$. Thus, the AFM and FM order should appear
in the large $U$ limit.

\begin{figure*}[t]
  \begin{center}
    \includegraphics[scale=0.4]{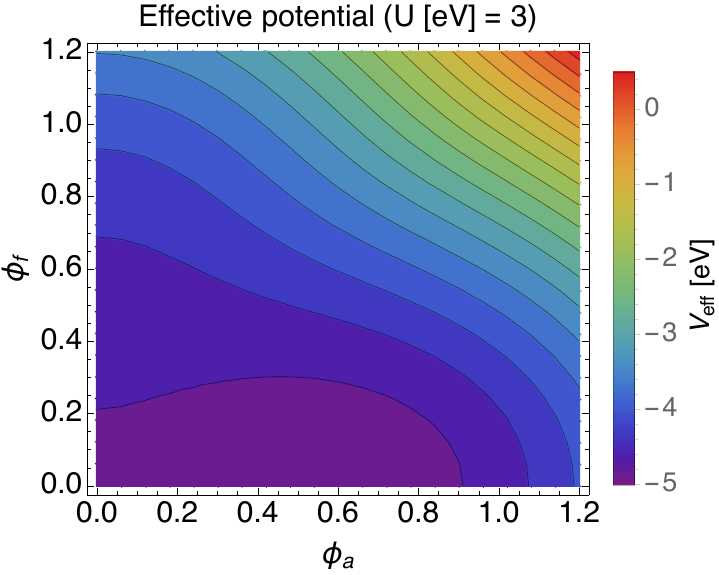}
    \includegraphics[scale=0.4]{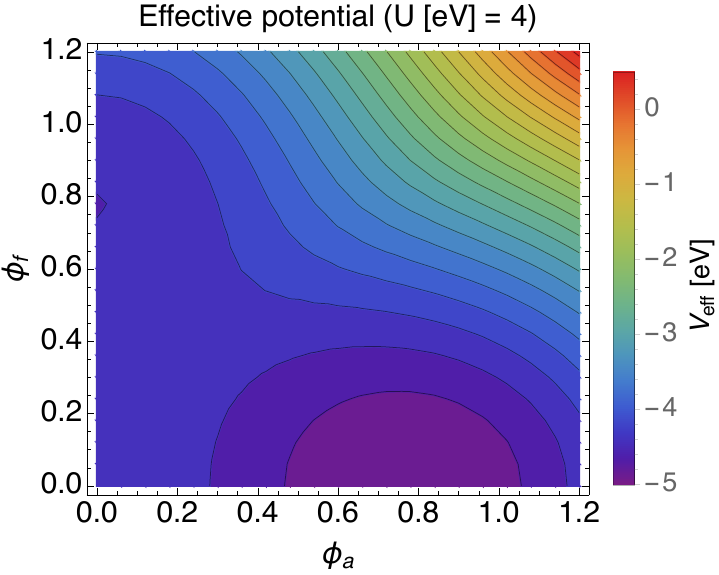}
    \includegraphics[scale=0.4]{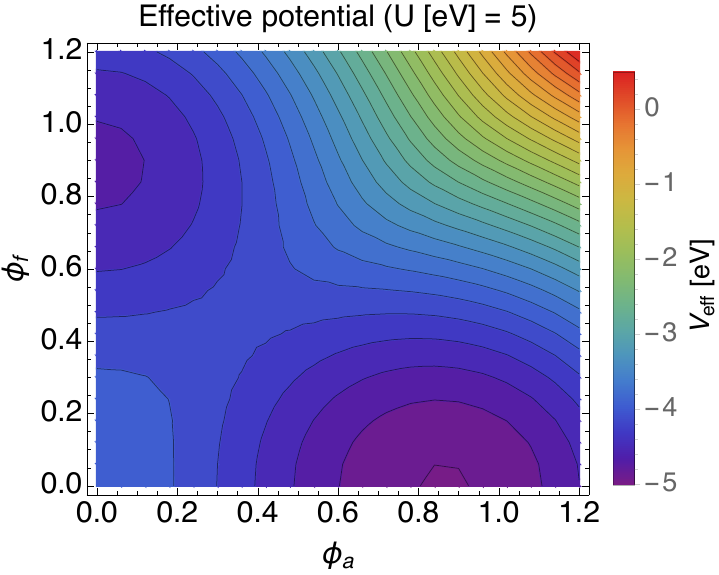}
    \includegraphics[scale=0.4]{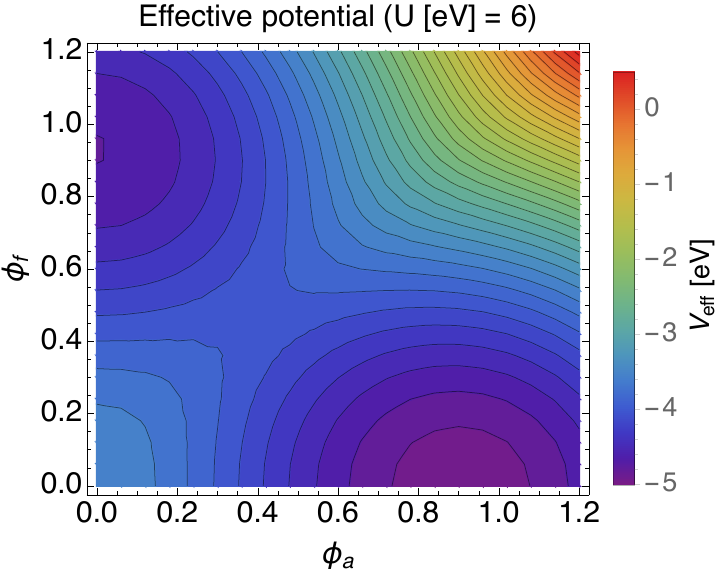}
    \includegraphics[scale=0.4]{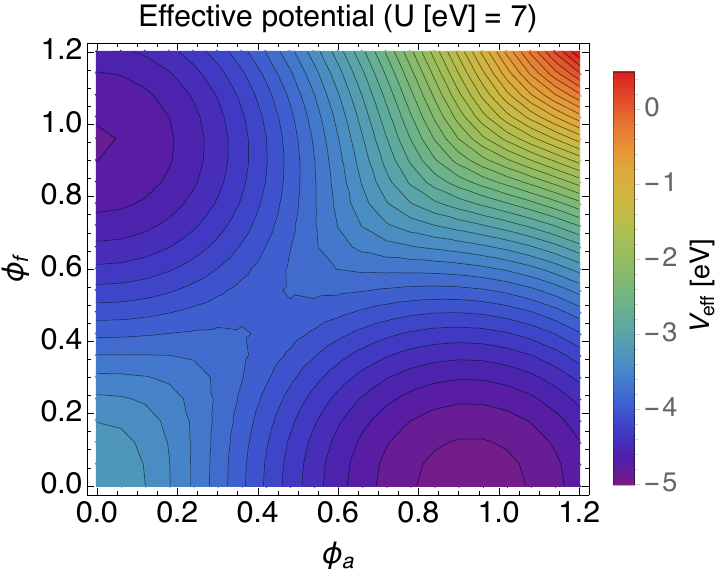}
    \includegraphics[scale=0.4]{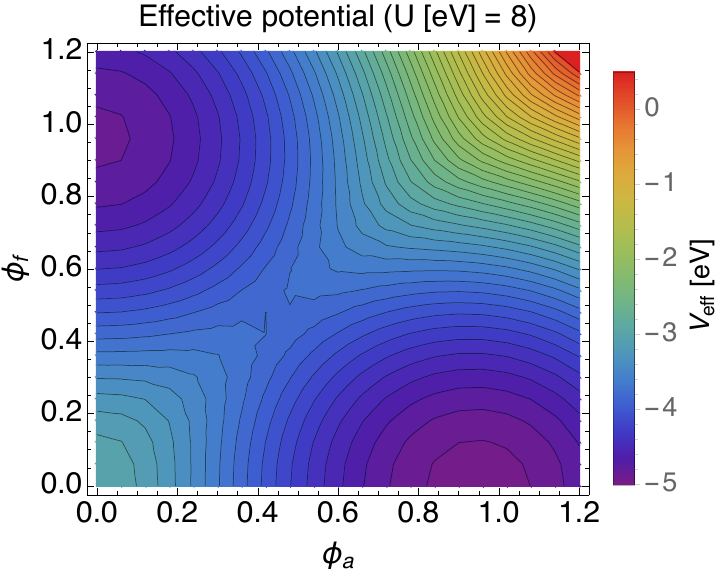}
  \end{center}
  \caption{Effective potential on $(\phi_a,\phi_f)$ plane at
    $T=0$. Each panel corresponds to different values of $U$. The
    other parameters are $A_1=A_2=1~{\rm eV}$, $B_1=B_2=-0.05~{\rm
      eV}$, and $M_0=0.01$~eV. Their definitions are given in
    Eq.\,\eqref{eq:d_a} and the below. }
  \label{fig:Veff}
\end{figure*}

\begin{figure*}[t]
  \begin{center}
    \includegraphics[scale=0.5]{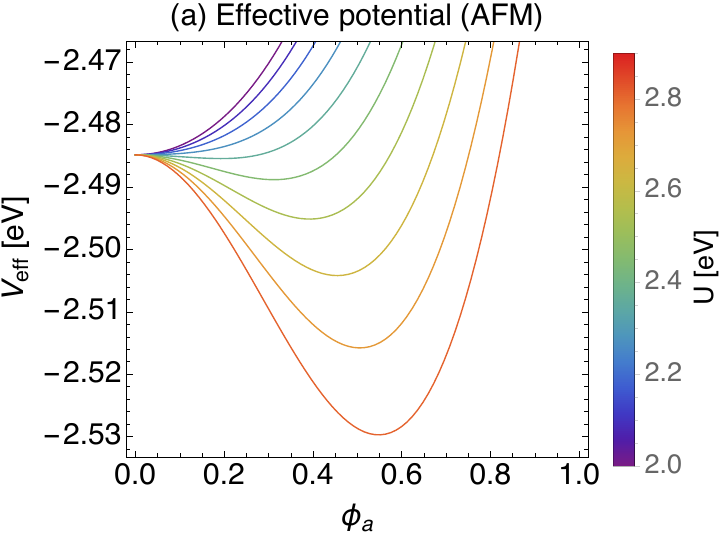}~~~~~
    \includegraphics[scale=0.5]{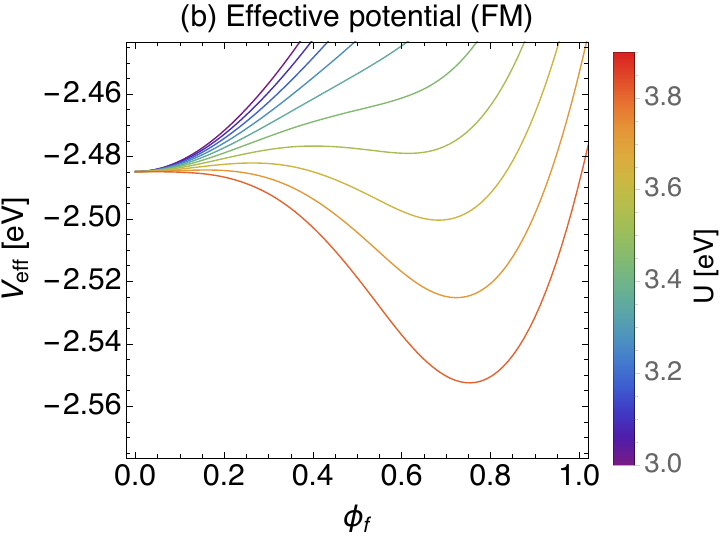}
  \end{center}
  \caption{
    (a) Effective potential under $\phi_f=0$ as function of $\phi_a$
    with various values of $U$ indicated in the color bar. (b) The same
    as (a) but under $\phi_a=0$ as function of $\phi_f$.  For both
    panels, the parameters $A_1, A_2, B_1, B_2$, and $M_0$ are the
    same as Fig.\,\ref{fig:Veff}.}
  \label{fig:phis+V}
\end{figure*}

\subsection{Numerical results}

Now we are ready to show the effective potential on $(\phi_a,\phi_f)$
plane. Fig.\,\ref{fig:Veff} shows the effective potential for a given
set of parameters. We take $T=0$ and the normal insulator (NI) phase
is considered. The TI phase can be calculated by taking $M_0<0$ with
the others being unchanged and we found the result merely changes. It
is seen the potential minima appear as $U$ increases.  As $U$ gets
large, the AFM order forms first since the condition for the AFM
order, i.e., \eqref{eq:PT_AFM}, is relatively easy to satisfy, and
eventually the FM order shows up.

To see the emergence of the magnetic orders, Fig.\,\ref{fig:phis+V}
shows the effective potential as function $\phi_a$ or $\phi_f$.  It
should be noted that $\phi_{a0}$ appears continuously from zero
meanwhile the non-zero $\phi_{f0}$ is obtained discontinuously. For
the FM order, it is seen that another vacuum appears as the $U$
becomes larger while the coefficient of the quadratic term is
positive. This is due to the negative $\phi_f^4$ term in
Eq.\,\eqref{eq:Veff_f}. Such a situation never happens for the AFM
since the quartic term is always positive. In fact, the phase
transition to the AFM order happens when the value of $U$ exceeds a
critical value $U_c^{\rm AFM}$ defined by
\begin{align}
  U_c^{\rm AFM}=\left(\frac{1}{2N}\sum_{\vb*{k}}\frac{1}{d_0}\right)^{-1}\,.
  \label{eq:U_cAFM}
\end{align}
In the present parameter set, i.e., $A_1=A_2=1$~eV, $B_1=B_2=-0.05$~eV,
and $M_0=0.01$~eV, we find $U_c^{\rm AFM}\sim 2~$eV.

Regarding to the ground state of the magnetic order, we have
numerically checked that the minimum of the effective potential is
obtained when the magnetic state is the AFM. This is true for other
sets of parameters. Therefore, the ground state of the system is the
AFM order. On the other hand, both minima for the AFM and FM orders
get degenerate as $U\to \infty$.  This can be understood
analytically. In the large $U$ limit, the stationary points
$(\phi_{a0},\phi_{f0})$ reduce to $(1,0)$ and $(0,1)$ and
$\epsilon_{1\vb*{k}}$, $\epsilon_{2\vb*{k}}\simeq U/2$. Thus both
minima approach to
\begin{align}
  V_{\rm eff}|_{T=0} \to
  \frac{U}{2} - \frac{1}{N}\sum_{\vb*{k}}2\times \frac{U}{2}
  =-\frac{U}{2}\,.
\end{align}
Therefore, when the magnetization is saturated, i.e.,
$\phi_{a0},\phi_{f0} \simeq 1$, both AFM and FM orders are expected to
be realized as quasi-degenerate vacua.

To summarize, we have seen the emergence of the AFM and FM orders by
computing the effective potential from the grand potential and found
that the AFM is the ground state. On the other hand, the AFM and FM
states are degenerate as $U$ gets larger.

\section{The amplitude mode and magnon under magnetic order}
\label{sec:excitations}

\begin{figure*}[t]
  \includegraphics[scale=0.5]{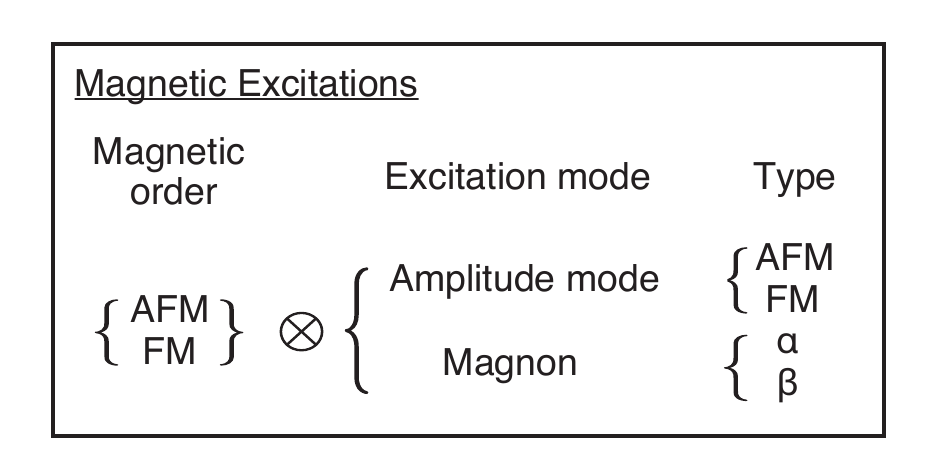}~~~~~~~~
  \includegraphics[scale=0.4]{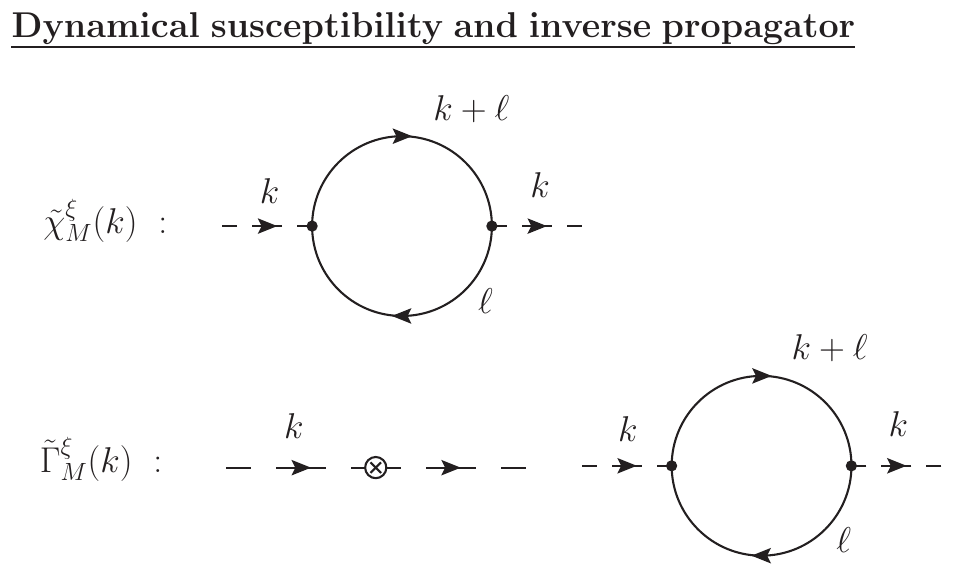}
  \caption{(Left) Classification of the magnetic
    excitations. Excitation modes are amplitude mode and magnon. The
    amplitude mode has `AFM-type' and `FM-type,' defined in
    Eqs.\,\eqref{eq:amp_modes1} and \eqref{eq:amp_modes2}, and the
    magnon has `$\alpha$-type' and `$\beta$-type,' defined in
    Eqs.\,\eqref{eq:mag_modes1} and \eqref{eq:mag_modes2}. Images of
    the amplitude mode and magnons are shown in Figs.\,\ref{fig:amp}
    and \ref{fig:mag}, respectively. (Right) Diagram corresponding to
    dynamical susceptibility $\tilde{\chi}^\xi_M(k)$ and the
    (dimensionless) inverse propagator $\tilde{\Gamma}^\xi_M(k)$.
    $k=(i\omega_n,\vb*{k})$ and $\ell=(i\omega_\ell,\vb*{\ell})$ are
    external and loop momenta, respectively.  Dashed line shows the
    magnetic excitations, amplitude mode or magnon, and solid line
    shows electron. In the definitions of $\tilde{\chi}_M^\xi$ and
    $\tilde{\Gamma}_M^\xi$, the external lines of the excitations are
    excluded. Blobs are vertices and a set of $\Gamma^5$,
    $\Gamma^{12}$, $\Gamma^\pm$, or $\Sigma^{\pm}$ enters depending on
    the excitations. Crossed blob stands for the tree-level vertex or
    tadpole.}
\label{fig:diagram}
\end{figure*}

In this section, we evaluate the gap and dispersion relation of the
magnetic excitations under the magnetic orders.  As seen in the
previous section, the possible magnetic orders are the AFM and FM and
the former is the ground state. To consider the excitations under both
magnetic environments, we take $(\phi_{ai},\phi_{fi})=(\phi_{a0},0)$
or $(0,\phi_{f0})$, where $\phi_{a0}$ and $\phi_{f0}$ are constants
that satisfy Eqs.\,\eqref{eq:phi_a0} and \eqref{eq:phi_f0},
respectively.  We will see that there are two magnetic excitations,
that are amplitude mode and magnon, and each of them has two
types. Fig.\,\ref{fig:diagram} (left panel) shows the classification
of the excitations. We will also discuss an excitation called `axion'
in the magnetic TIs.

\subsection{Dynamical susceptibility and inverse propagator}

For the study we introduce a dynamical susceptibility, which is defined
by
\begin{align}
  \chi_M(k;O_1,O_2)
 =-\frac{1}{\beta N}\sum_{i\omega_\ell,\vb*{\ell}}
  {\rm Tr}[\tilde{G}_M(\ell)O_1\tilde{G}_M(\ell+k)O_2]\,,
  \label{eq:def_chi}
\end{align}
where $\tilde{G}_M$ is the Green's function of the electron under the
global magnetic order, AFM or FM, in the wavenumber space, which will
be defined soon. The index $M=A$ and $F$ stands for the magnetic
order, i.e., AFM and FM, respectively.  $O_1$ and $O_2$ are four by
four matrices, such as Gamma matrices. For the arguments of the
susceptibility and the Green's function, we write $k$ and $\ell$ as
$k=(i\omega_n,\vb*{k})$ and $\ell=(i\omega_\ell,\vb*{\ell})$. Instead
of $\chi_M$, we will use a dynamical susceptibility that is symmetric
with respect to $O_1$ and $O_2$:
\begin{align}
  \tilde{\chi}^{\xi}_M(k)\equiv 
  \frac{1}{2}\left\{\chi_M(k;O_1,O_2)+\chi_M(k;O_2,O_1)\right\}\,,
\end{align}
Here $\xi$ is a label for a specific set of operators $O_1$ and $O_2$,
which will be given explicitly in this section. 

With the dynamical susceptibility, we will see that
$\tilde{\Gamma}^\xi_M(k)$ defined by
\begin{align}
  \tilde{\Gamma}^\xi_M(k)\equiv 1-\frac{U}{4}\tilde{\chi}^\xi_M(k)\,,
  \label{eq:tilde_Gamma^xi_M}
\end{align}
plays the role of the inverse propagator, which is normalized to be
dimensionless, of the excitations at the one-loop
level. Fig.\,\ref{fig:diagram} (right panel) shows diagrams that
correspond to the dynamical susceptibility $\tilde{\chi}_M^\xi$ and
the inverse propagator $\tilde{\Gamma}^\xi_M$.

\subsection{The amplitude mode}
\label{sec:amp}

\begin{figure*}[t]
  \begin{center}
    \includegraphics[scale=0.35]{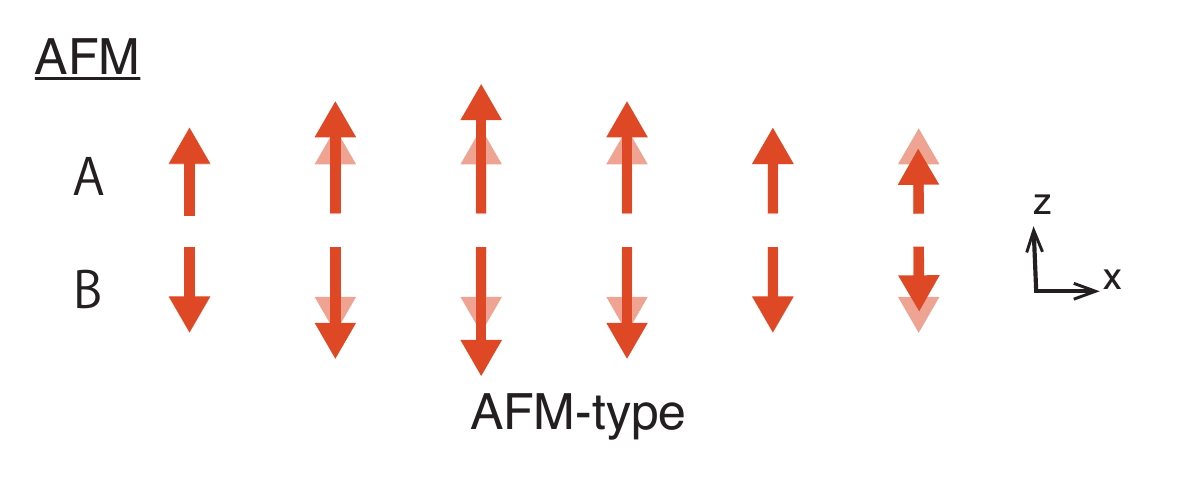}~~~
    \includegraphics[scale=0.35]{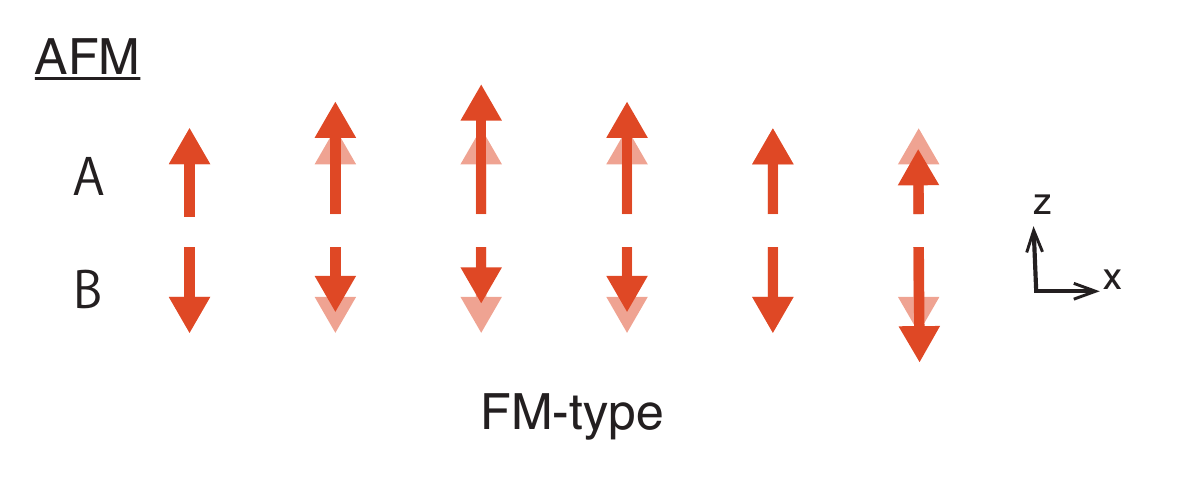}
    \includegraphics[scale=0.35]{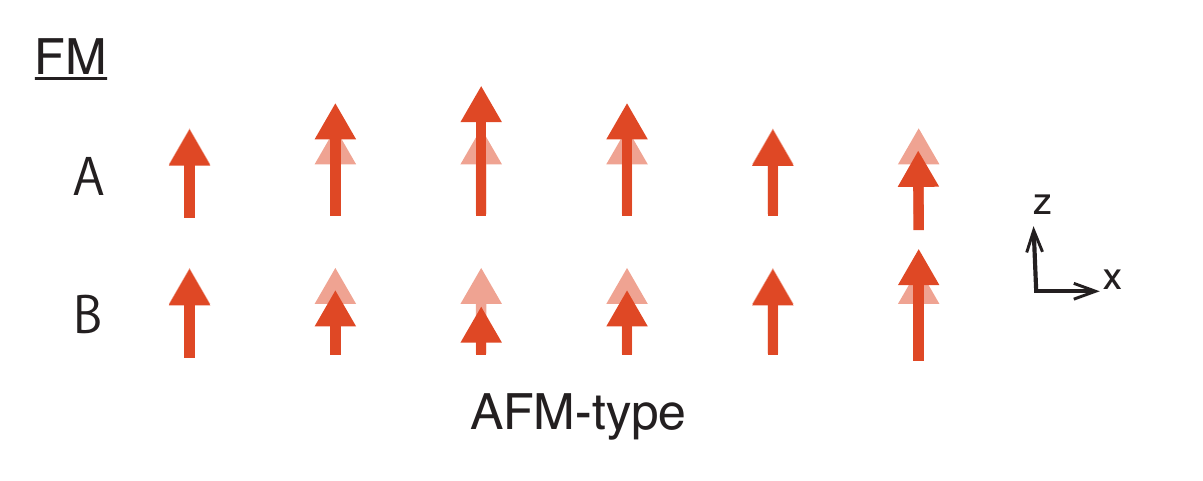}~~~
    \includegraphics[scale=0.35]{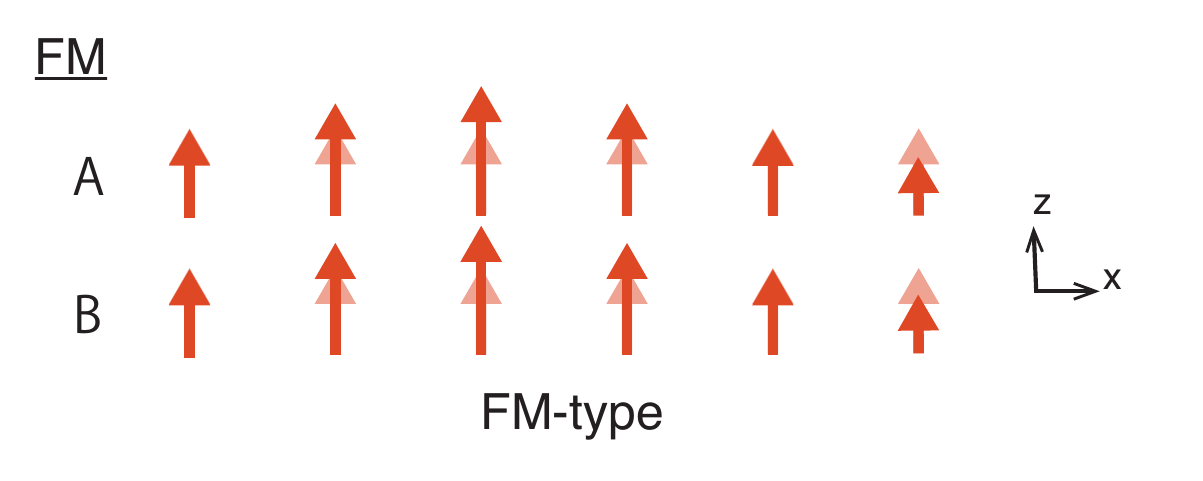}
  \end{center}
  \caption{Image of the amplitude mode under the AFM (top) and FM
    (bottom) orders.  The AFM-type (left) and FM-type (right) is
    shown. The plot is an example of the excitation with a wavenumber
    $\vb*{k}=(\pi/4,0,0)$. }
  \label{fig:amp}
\end{figure*}

To discuss the amplitude mode, we take
\begin{align}
  \phi_{ai}&=\phi_{a0}+\delta \phi_{ai}\,,
  \label{eq:amp_modes1} \\
  \phi_{fi}&=\phi_{f0}+\delta \phi_{fi}\,.
  \label{eq:amp_modes2}
\end{align}
$\delta\phi_{ai}$ and $\delta \phi_{fi}$ are two independent
fluctuations of the AFM and FM order parameters, respectively, and we
call them `AFM-type' and `FM-type.' The example of the fluctuations
are shown in Fig.\,\ref{fig:amp}.  One can imagine that there are four
excitations by looking at the effective potential in
Fig.\,\ref{fig:Veff}; the AFM- and FM-type amplitude modes under the
AFM correspond to the fluctuations around
$(\phi_a,\phi_f)=(\phi_{a0},0)$ in $\phi_a$ and $\phi_f$ directions,
respectively, and a similar interpretation applies to the fluctuations
around $(\phi_a,\phi_f)=(0,\phi_{f0})$.  From the figure we naively
expect all the amplitude modes to be stable. This intuition, however,
needs to be modified, which will be discussed below.

With Eqs.\,\eqref{eq:amp_modes1} and \eqref{eq:amp_modes2}, we expand
the second term of Eq.\,\eqref{eq:Seff_orig} as\footnote{Similar
expansion is done in
Refs.\,\cite{Sekine:2015eaa,Sekine_2021,Schutte-Engel:2021bqm} to
derive `axion' mass, meanwhile the first term of
Eq.\,\eqref{eq:H_U_1a}, i.e., $U\phi^2_{ai}/2$ that will give rise to
a term $U\delta\phi^2_{ai}/2$ in Eq.\,\eqref{eq:Samp_pre}, is missing
at the beginning. When this term is missing, then the resultant `mass
term' gets the opposite sign, leading to the tachyonic
instability. The authors, however, claim that they obtain the same
result as one in Ref.\,\cite{Li:2009tca} though their calculation is
inconsistent. See also discussion below Eq.\,\eqref{eq:m_a}.}
\begin{align}
  -\ln \det (\partial_\tau +{\cal H})
  &=-{\rm Tr}\ln(\partial_\tau+{\cal H})
  \nonumber \\
  &=-{\rm Tr}\ln (-G_M^{-1})+
  \sum_{n=1}^\infty \frac{1}{n}{\rm Tr}(G_M\delta {\cal H})^n\,,
\end{align}
by decomposing ${\cal H}={\cal H}^{\rm MTI}+\delta {\cal H}$
where
\begin{align}
  {\cal H}^{\rm MTI} &= {\cal H^{\rm TI}}
  +(U/2)(\phi_{a0}\Gamma^5+\phi_{f0}\Gamma^{12})\,,
  \\
  \delta {\cal H} &=
  (U/2)(\delta \phi_{ai}\Gamma^5+\delta \phi_{fi}\Gamma^{12})\,.
\end{align}
Then the Green's function is given by $G_M^{-1}=-\partial_\tau-{\cal
  H^{\rm MTI}}$ where $\phi_{f0}=0$ and $\phi_{a0}=0$ for $M=A$ and
$F$, respectively.  Using the definition of the Fourier expansion of
the field and the Green's function given in
Appendix~\ref{app:Fourier_ex}, the effective action at $\order{\delta
  \phi_{ai}^2,\delta \phi_{fi}^2}$ is
\begin{align}
  S_{\rm eff} &\supset \int_0^\beta d\tau \sum_i \sum_{\xi=a,f}
  \frac{U}{2}\delta \phi_{\xi i}^2
  +\frac{1}{2}{\rm Tr}(G\delta {\cal H})^2
  \label{eq:Samp_pre}\\ 
  &=\frac{U}{2}\sum_{i\omega_n,\vb*{k}}\sum_{\xi=a,f}
  \tilde{\delta \phi_\xi}(k)\tilde{\delta \phi_\xi}(-k)
   \tilde{\Gamma}^\xi_M(k)\,,
  \label{eq:Samp}
\end{align}
where $\tilde{\Gamma}^\xi_M$ is given in Eq.\,\eqref{eq:tilde_Gamma^xi_M}
and 
\begin{align}
  \tilde{\chi}_M^a(k)&=\chi_M(k;\Gamma^5,\Gamma^5)\,,
  \\
  \tilde{\chi}_M^f(k)&=\chi_M(k;\Gamma^{12},\Gamma^{12})\,.
\end{align}
The first term of $\tilde{\Gamma}^\xi_M(k)$ comes from the first term
in Eq.\,\eqref{eq:Samp_pre}, which is the tree-level contribution mass
term and the second one corresponds to the loop
diagram. $\tilde{\Gamma}^\xi_M(k)$ plays a role of an inverse
propagator of the excitations at the one-loop level while it is
normalized to dimensionless in our notation. Thus
$1/\tilde{\Gamma}^\xi_M(k)$ plays a role of the connected two-point
Green's function that is given by summing over the infinite series of
the one-loop diagram.  $\tilde{\chi}^\xi_M(k)$ can be evaluated in the
basis where the Green's functions are diagonalized. The calculation
is shown in Appendix~\ref{app:chi}.  From Eq.\,\eqref{eq:Samp}, the
gap and dispersion relation $\omega =\omega(\vb*{k})$ of the amplitude
mode is given by solving
\begin{align}
  \tilde{\Gamma}^\xi_M(i\omega_n=\omega+i\delta,\vb*{k})
  = 0\,,
  \label{eq:gap_eq}
\end{align}
where $\delta \ll 1$.

\begin{figure*}[t]
  \begin{center}
    \includegraphics[scale=0.5]{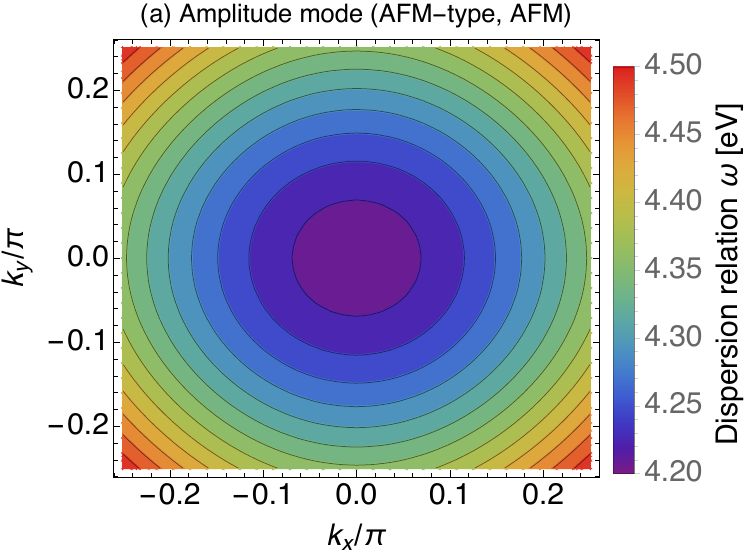}~~~~~
    \includegraphics[scale=0.5]{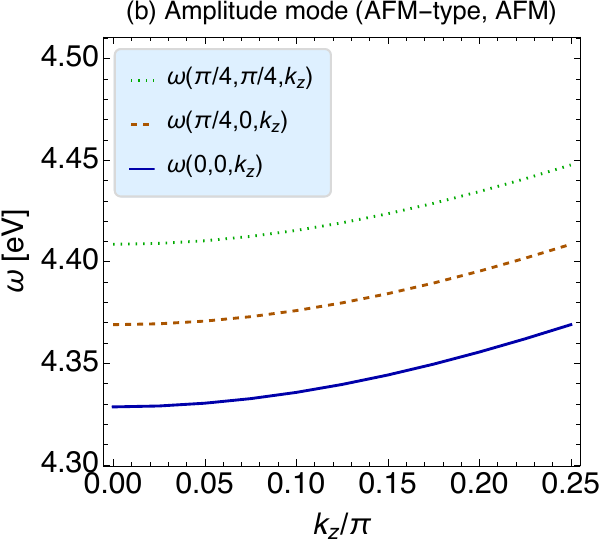}
  \end{center}
  \caption{Dispersion relation for the AFM-type amplitude mode under
    the AFM.  (a) Contour of $\omega(k_x,k_y,0)$. (b)
    $\omega(\vb*{k})$ as function of $k_z$ for various values of
    $k_x,k_y$.  We take $U=5$~eV and the other parameters are the same
    as Fig.\,\ref{fig:Veff}.}
  \label{fig:omg_AFM-amp_AFM}
\end{figure*}

Before the numerical study, we comment on the `axionic' excitation in
the magnetic TIs.  We claim that the AFM-type amplitude mode under the
AFM order corresponds to the quasi-particle `axion,' regardless of the
topological nature of the insulators. Namely, the `axion' corresponds
to the AFM-type amplitude mode in both TI and NI phases.\footnote{The
existence of the dynamical `axion' in NIs with the AFM state is
already pointed out in Ref.\,\cite{Ooguri:2011aa}. The dynamical
`axion' is also studied under the FM order in
Ref.\,\cite{Wang:2015hhf,Ishiwata:2022mzq}.}  First of all, the gap of
the AFM-type amplitude mode under the AFM order, obtained by
Eq.\,\eqref{eq:gap_eq}, corresponds to the mass of the `axion'. This
fact is understood from the fact that $\tilde{\Gamma}^m_M$ is the
inverse propagator of the excitation and that the pole of the
propagator gives the physical mass.  To derive the effective action
for the `axion,' let us take the zero temperature limit. First, the
mass $m_a$ of `axion' is given by
\begin{align}
  \tilde{\Gamma}^a_A(i\omega_n=m_a,\vb*{0})
  = 0\,.
\end{align}
It is easy to find a solution for the equation by using an analytic
expression for $\tilde{\chi}^a_A$ given Eq.\,\eqref{eq:chi_A^a} of
Appendix~\ref{app:chi_AFM}:
\begin{align}
  m_a=U\phi_{a0}\,(=2d_5)\,,
  \label{eq:m_a}
\end{align}
for $\phi_{a0}\neq 0$.\footnote{The `axion' mass can be obtained by
solving $\tilde{\Gamma}^a_A(i\omega_n=m_a,\vb*{0}) = 0$ numerically
for $\phi_{a0}=0$ case. }  Then, expanding $\tilde{\Gamma}^a_A$ at
$i\omega_n=m_a$ and $\vb*{k}=\vb*{0}$ and taking a continuum limit for
the space, we get the action of the `axion' in the Minkowski spacetime
as
\begin{align}
  S_a =
  \left(\frac{U}{2}\right)^2 J \int d^4x~
  \delta \phi_a\left[-\partial_t^2+v_i^2\partial^2_i
    -m_a^2 \right]\delta \phi_a\,,
  \label{eq:S_axion}
\end{align}
where an index $i$ is summed from  1 to 3 and 
\begin{align}
  J=\int \frac{d^3\ell}{(2\pi)^3}\frac{1}{4dd_0^2}
  \,,
  \label{eq:J}
\end{align}
Here $J$ and $v_i$ are the stiffness and velocity, respectively. See
Appendix~\ref{app:L} for the derivation of the action. The results of
the mass and stiffness given in Eqs.\,\eqref{eq:m_a} and \eqref{eq:J}
are different from the previous
results~\cite{Ishiwata:2021qgd}\footnote{$J$ and $Jm_a^2$ in
Refs.\,\cite{Li:2009tca,Schutte-Engel:2021bqm,Sekine:2015eaa,Sekine_2021}
are smaller than those in Ref.\,\cite{Ishiwata:2021qgd} by a factor of
4, which was already pointed out in Ref.\,\cite{Ishiwata:2021qgd}.}
\begin{align}
  J^{\rm old}&=\int \frac{d^3\ell}{(2\pi)^3}\frac{d^2_0}{4d^5}\,,
  \label{eq:J_old}
  \\
  J^{\rm old}(m_a^{\rm old})^2&
  =\int \frac{d^3\ell}{(2\pi)^3}\frac{d_5^2}{d^3}\,.
  \label{eq:Jma_old}
\end{align}
The reason is that the previous results are given by expanding
$\tilde{\Gamma}^a_A(i\omega_n,\vb*{0})$ at $i\omega_n=0$, not
$i\omega_n=m_a$. This is based on the assumption that the expansion
around $i\omega_n=0$ is a good approximation. However, the
quantitative argument regarding the validity of the expansion around
$i\omega_n=0$ has not been clarified explicitly in the previous
works.

Compared to the previous calculation, we have two improvements.
First, the mass and stiffness given in the present work are more
accurate. This is because we do not use perturbative expansion to
derive the mass and that the stiffness is derived by the derivative of
$\tilde{\Gamma}^a_A$ with respect to $(i\omega_n)^2$ at the
mass. Second, it is possible to check whether the gap (or mass)
obtained at $\vb*{k}=0$ is the minimum. Namely, there is a possibility
that a smaller gap is obtained for a nonzero $\vb*{k}$.  In such a
case, we must expand $\tilde{\Gamma}^a_A$ at the wavenumber to derive
the effective action for the lower-energy excitation.  We will check
this possibility numerically below.

By the use of the effective action, we can evaluate the validity
condition of the perturbative expansion of the past works. As a
result, we found that the expansion is valid for $m_a\lesssim
\order{{\rm eV}}$. The details are given in Appendix~\ref{app:L}.

Fig.\,\ref{fig:omg_AFM-amp_AFM} shows the dispersion relation of the
AFM-type amplitude mode under the AFM order. We take $U=5$ and the
other parameters are the same as Fig.\,\ref{fig:Veff}. We found that
$\vb*{k}=\vb*{0}$ is the stable minimum. Here `stable' means that the
curvature of the dispersion relation $\omega=\omega(\vb*{k})$ is
positive in all $\vb*{k}$ directions around $\vb*{k}=\vb*{0}$. In
other word, $v_i^2$ are positive for all $i=1$\,--\,3. This can be
seen explicitly in the figure. We have checked that there are no
smaller gaps, thus the gap is surely given by $m_a=U\phi_{a0}\sim
\order{\rm eV}$.

The situation is different for the other amplitude modes. As an
example, let us discuss the FM-type amplitude mode with the AFM
order. The dynamical susceptibility of the excitation is analytically
obtained by (see Appendix~\ref{app:chi_AFM})
\begin{align}
  \tilde{\chi}_A^{f}(\omega+i\delta,\vb*{k})
  =
  \frac{2}{N}\sum_{\vb*{\ell}}
  \frac{df-\sum_{a=1}^5d_af_a+2(d_1f_1+d_2f_2)}{df(d+f)}
  \frac{1}{1-(\omega^2+i\delta)/(d+f)^2}\,,
\end{align}
where $d_a$ and $f_a$ $(a=1$\,--\,$4)$ depend on the wavenumber as
$d_a=d_a(\vb*{\ell})$, $f_a=d_a(\vb*{\ell}+\vb*{k})$, $f_5=d_5$, and
$f=(\sum_{a=1}^5f_af_a)^{1/2}$.  When $\vb*{k}=0$, for instance, we
found that the gap is found to satisfy $\omega(\vb*{0})>\omega_c$,
where $\omega_c={\rm Min} (2d)$. In addition, the imaginary part of
$\tilde{\chi}^f_A$ is enhanced. This indicates that the amplitude mode
tends to dissipate to an electron-hole pair. One can understand the
interpretation of the imaginary part from a corresponding diagram to
the dynamical susceptibility, shown in Fig.\,\ref{fig:diagram}.
Similar behavior is found for nonzero $\vb*{k}$.  Therefore, the
excitation is not stable and dissipates even if it is created. We have
also found no stable configuration for both AFM- and FM-type amplitude
modes under the FM order. In addition, the expansion for $\omega$ is
inappropriate since $\omega/(d+f)$ can be larger than unity.
Therefore, the effective description of the form of
Eq.\,\eqref{eq:S_axion} is invalid for those excitations.

\subsection{The magnon}
\label{sec:magnon}

\begin{figure*}[t]
  \begin{center}
    \includegraphics[scale=0.35]{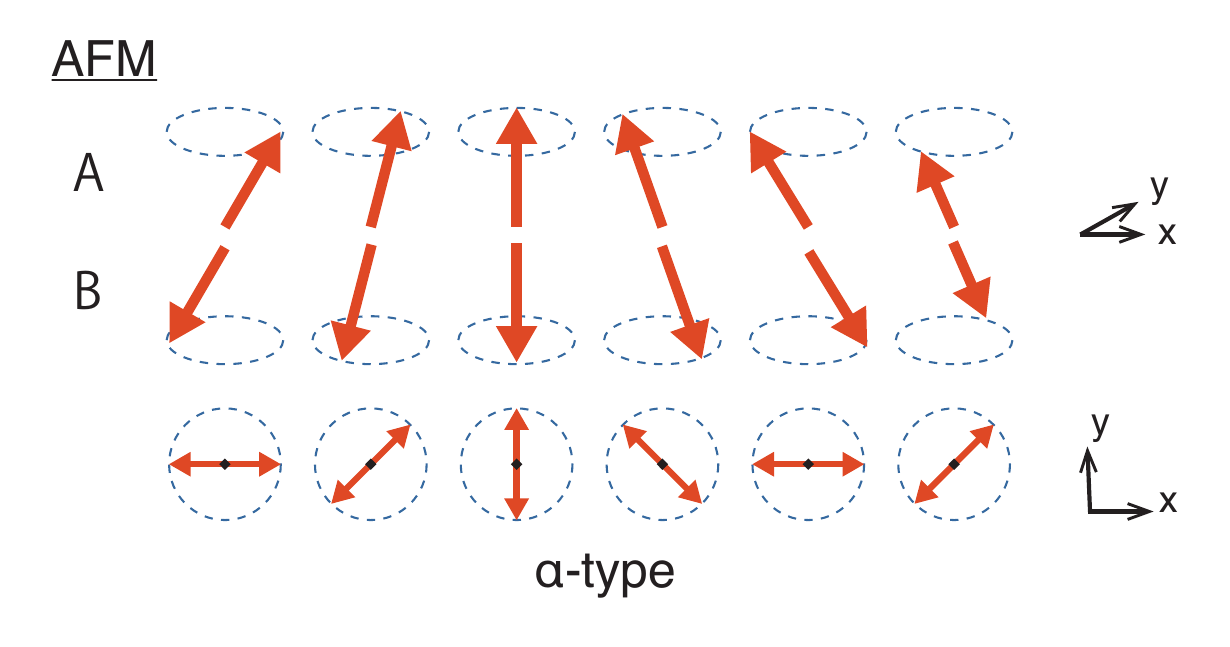}~~~~
    \includegraphics[scale=0.35]{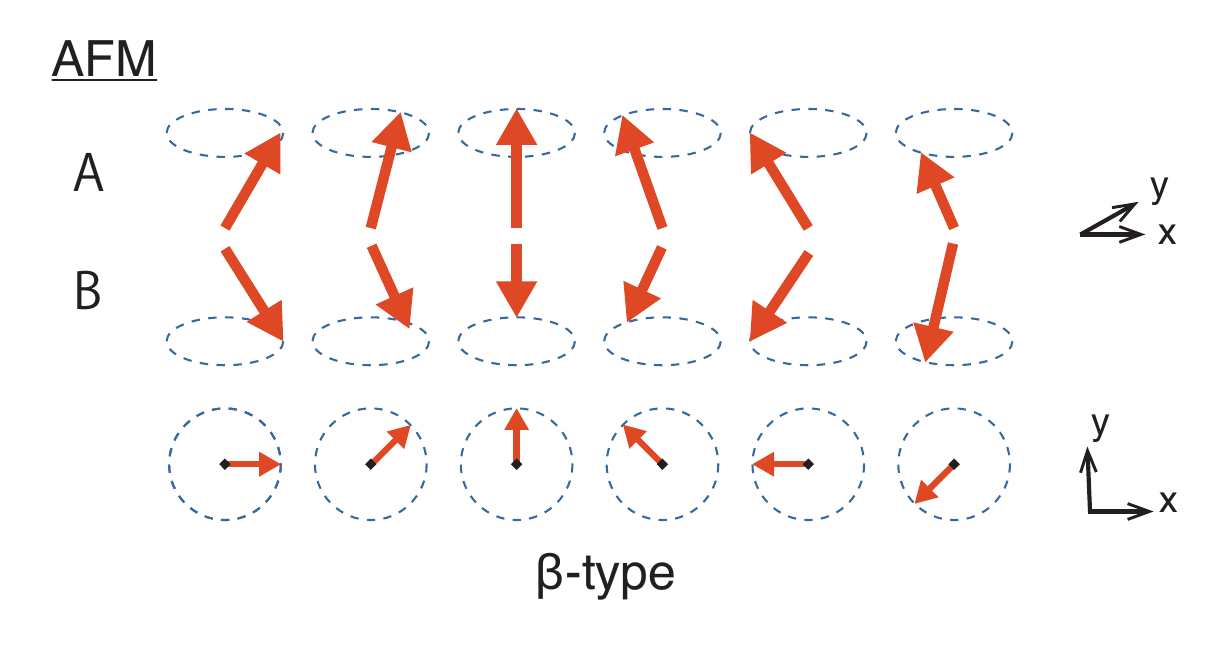}
    \includegraphics[scale=0.35]{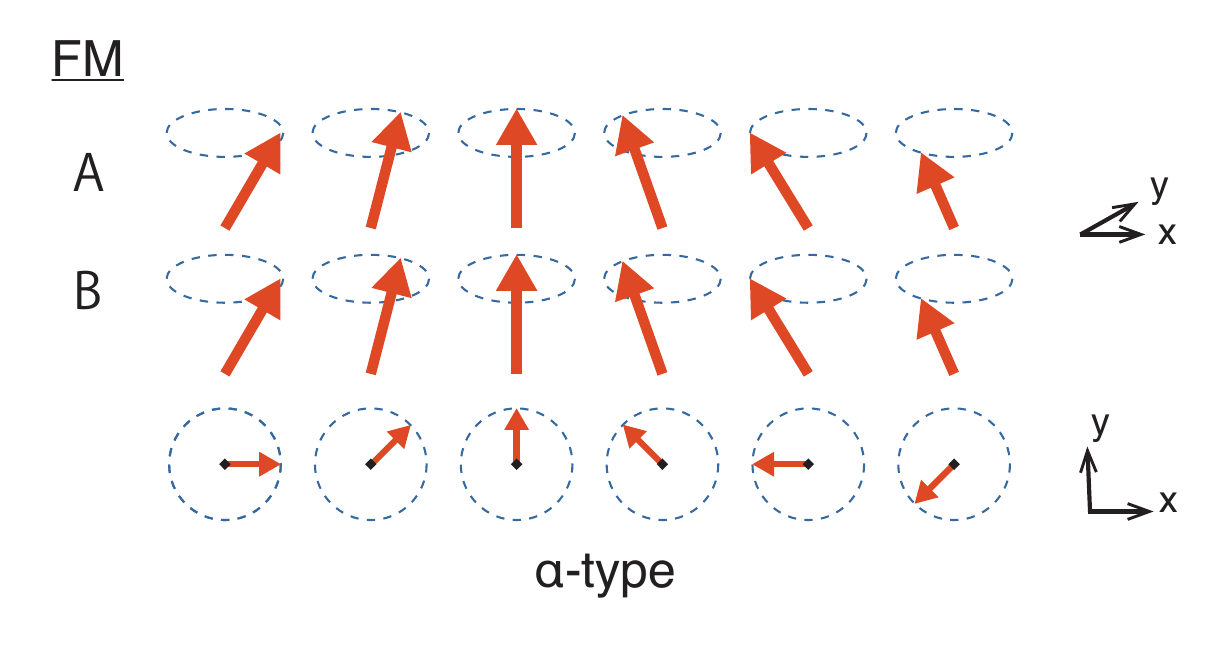}~~~~
    \includegraphics[scale=0.35]{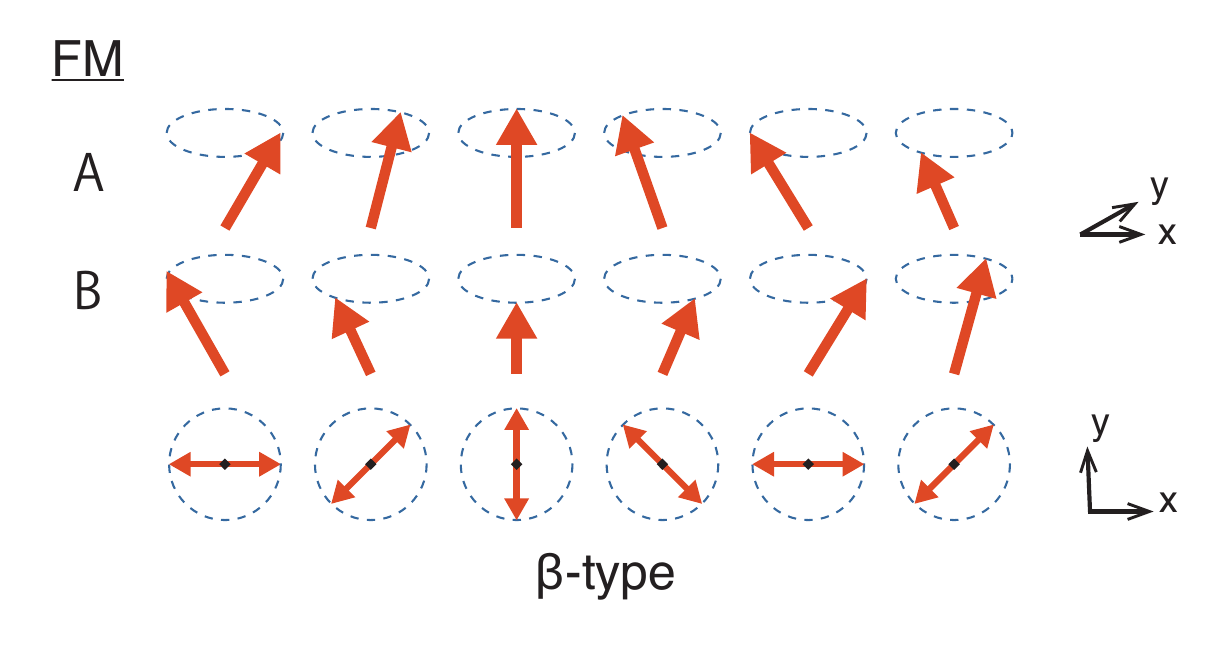}
  \end{center}
  \caption{Image of the magnon under the AFM (top) and FM (bottom)
    orders.  The plot is an example of the excitation with a
    wavenumber $\vb*{k}=(\pi/4,0,0)$. }
  \label{fig:mag}
\end{figure*}

To describe the magnon, we need to rewrite the Hubbard term in an
isotropic form. Namely, the Stratonovich-Hubbard transformation gives
\begin{align}
  H_U \to
  &\frac{U}{4}\sum_{I,i}\left[
    \phi^2_I+2\phi_I
    c^\dagger_{Ii}\vec{n}_{Ii}\vdot \vec{\sigma}c_{Ii}
    \right]\,,
  \label{eq:H_U_2} 
\end{align}
where $\vec{n}_{Ii}=(n_{Ii}^x,n_{Ii}^y,n_{Ii}^z)$ that satisfies
$|\vec{n}_{Ii}|=1$. Under the AFM, $\phi_{Ai}=-\phi_{Bi}=\phi_{a0}$,
while $\phi_{Ai}=\phi_{Bi}=\phi_{f0}$ for the FM.  Recalling that we
are interested in the case where $z$ direction is the easy axis, the
fluctuation of $n_{Ii}^x,n_{Ii}^y$ corresponds to the magnon. For
later analysis we introduce $\vec{n}_{\alpha i}$ and $\vec{n}_{\beta
  i}$ as
\begin{align}
  \vec{n}_{\alpha i}\equiv (\vec{n}_{Ai}+ \vec{n}_{Bi})/2\,,
  \\
  \vec{n}_{\beta i}\equiv (\vec{n}_{Ai}- \vec{n}_{Bi})/2\,.
\end{align}
The labels $\alpha$ and $\beta$ stand for two different types of
magnon excitations, which we call `$\alpha$-type' and `$\beta$-type,'
respectively.  We will formulate each excitation under both the AFM
and FM orders below.

First, let us discuss the magnon under the AFM order.  $\vec{n}_A$ and
$\vec{n}_B$ are parallel for the $\alpha$-type, meanwhile rotation
phases of $\vec{n}_A$ and $\vec{n}_B$ coincide for the
$\beta$-type. See Fig.\,\ref{fig:mag} (top panel) for an example of
the $\alpha$- and $\beta$-types under the AFM. The interaction
Hamiltonian gives
\begin{align}
  H_U \supset
  \frac{U\phi_{a0}}{2}\sum_{i}c^\dagger_i
  \Bigl[
    &\Gamma^5n_{\alpha i }^z
    +\Gamma^1n_{\alpha i}^x+\Gamma^2n_{\alpha i}^y
    \nonumber \\
    +&\Gamma^{12}n_{\beta i}^z+\Gamma^{25}n_{\beta i}^x+\Gamma^{51}n_{\beta i}^y
    \Bigr]c_i\,,
  \label{eq:H_U_magnon_AFM}
\end{align}
where
\begin{align}
  \Gamma^{25}=\mqty(\sigma^1 &0 \\ 0 &\sigma^1)\,,~
  \Gamma^{51}=\mqty(\sigma^2 &0 \\ 0 &\sigma^2)\,.
\end{align}
Assuming that $n^x_{\alpha i},n^y_{\alpha i},n^x_{\beta i},n^y_{\beta
  i},\ll 1$ the Hamiltonian reduces to ${\cal H}={\cal H}^{\rm
  MTI}+\delta {\cal H}$ where
\begin{align}
  \delta {\cal H} =
  \frac{U\phi_{a0}}{2}\Bigl[
    &-\frac{1}{2}\Gamma^5\{(n^x_{\alpha i})^2+(n^y_{\alpha i})^2
    +(n^x_{\beta i})^2+(n^y_{\beta  i})^2\}
    +\Gamma^1n^x_{\alpha  i}+\Gamma^2n^y_{\alpha  i}
    \nonumber \\
    &-\Gamma^{12}(n^x_{\alpha  i}n^x_{\beta  i}+n^y_{\alpha i}n^y_{\beta  i})
    +\Gamma^{25}n^x_{\beta  i}+\Gamma^{51}n^y_{\beta  i}
    \Bigr]\,.
\end{align}
The term proportional to $\Gamma^{12}$ does not contribute to AFM
order and we will ignore it hereafter. Now it is convenient to introduce
\begin{align}
  n^\pm_{\alpha i} \equiv (n^x_{\alpha i}\pm i n^y_{\alpha i})/\sqrt{2}\,,
  \label{eq:mag_modes1}\\
  n^\pm_{\beta i} \equiv  (n^x_{\beta i}\pm i n^y_{\beta i})/\sqrt{2}\,.
  \label{eq:mag_modes2}
\end{align}

\noindent Using them, $\delta {\cal H}$ is written as
\begin{align}
  \delta {\cal H} =
  \frac{U\phi_{a0}}{2}\Bigl[
    &-\Gamma^5(n^+_{\alpha i}n^-_{\alpha i}+
    n^+_{\beta i}n^-_{\beta i})
    \nonumber \\
    &+\Gamma^+n^-_{\alpha  i}+\Gamma^-n^+_{\alpha  i}
    +\Sigma^+n^-_{\beta  i}+\Sigma^-n^+_{\beta  i}
    \Bigl]\,,
  \label{eq:dH_mag_AFM}
\end{align}
where
\begin{align}
  \Gamma^\pm&= \frac{1}{\sqrt{2}}(\Gamma^1\pm i \Gamma^2)
  =\sqrt{2}\mqty(\sigma^\pm & 0 \\ 0 & -\sigma^\pm)~\,,
   \\
   \Sigma^\pm&= \frac{1}{\sqrt{2}}(\Gamma^{25}\pm i \Gamma^{51})
    =\sqrt{2}\mqty(\sigma^\pm & 0 \\ 0 & \sigma^\pm)~\,.
\end{align}
Now we derive the effective action in a similar way as in
Sec.\,\ref{sec:amp}. Computing the terms up to
$\order{(n^\pm_{\alpha i})^2,(n^\pm_{\beta i})^2}$, the effective action
for the excitations is
\begin{align}
  S_{\rm eff} \supset
  \frac{U\phi_{a0}^2}{2}\sum_{i\omega_n,\vb*{k}}\sum_{\xi=\alpha,\beta }
  \tilde{n}_{\xi}^+(k)\tilde{n}_{\xi}^-(-k)\tilde{\Gamma}_A^\xi(k)\,,
  \label{eq:Seff_A_mag}
\end{align}
where $\tilde{\Gamma}^\xi_M$ is given in
Eq.\,\eqref{eq:tilde_Gamma^xi_M} and $\tilde{\chi}_A^\xi(k)$
($\xi=\alpha,\beta$) are given by
\begin{align}
   \tilde{\chi}_A^{\alpha}(k)&=
  \frac{1}{2}\{\chi_A(k;\Gamma^+,\Gamma^-)
  +\chi_A(k;\Gamma^-,\Gamma^+)\}\,,\\
  \tilde{\chi}_A^{\beta}(k)&=
  \frac{1}{2}\{\chi_A(k;\Sigma^+,\Sigma^-)
  +  \chi_A(k;\Sigma^-,\Sigma^+)\}\,.
\end{align}
Here the first term of $\tilde{\Gamma}_A^\xi$ is given by the term
proportional to $\Gamma^5$ in
Eq.\,\eqref{eq:dH_mag_AFM}. Diagrammatically it is a tadpole and we
have found that it can be rewritten by using
Eq.\,\eqref{eq:phi_a0}. As a result, it turns out to have the same
structure as the amplitude modes.

\begin{figure*}[t]
  \begin{center}
    \includegraphics[scale=0.5]{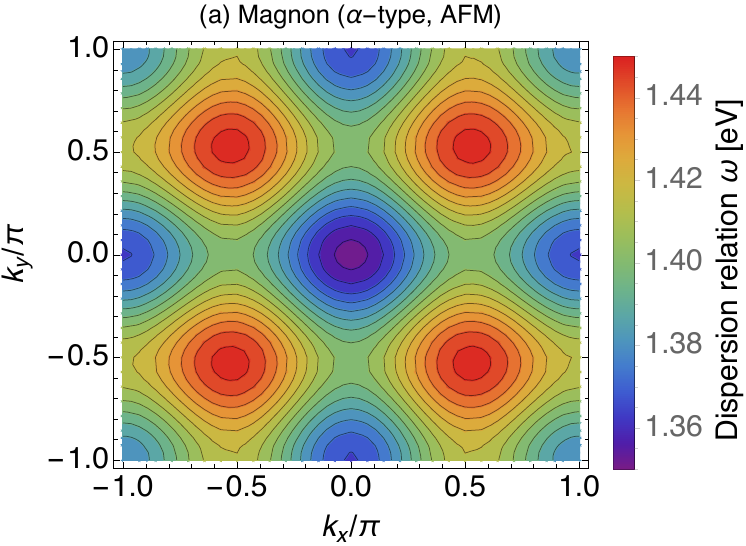}~~~~~
    \includegraphics[scale=0.5]{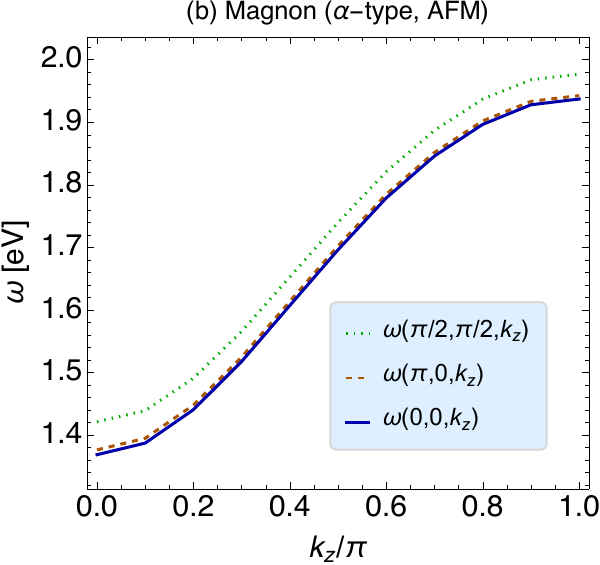}
  \end{center}
  \caption{Dispersion relation for the $\alpha$-type magnon under the AFM.
    (a) Contour of $\omega(k_x,k_y,0)$. (b) $\omega(\vb*{k})$
    as function of $k_z$ for various values of $k_x,k_y$.  We take
    $U=5$~eV and the other parameters are the same as
    Fig.\,\ref{fig:Veff}.}
  \label{fig:omg_AFM-mag_AFM}
\end{figure*}

We repeat the argument for the FM order. In this case, we take
$\phi_{Ai}=\phi_{Bi}=\phi_{f0}/2$ to give the Hamiltonian $H_U$
\begin{align}
  H_U \supset
  \frac{U\phi_{f0}}{2}\sum_{i}c^\dagger_i
  \Bigl[
    &\Gamma^{12}n_{\alpha i}^z+\Gamma^{25}n_{\alpha i}^x+\Gamma^{51}n_{\alpha i}^y
    \nonumber \\
    +&\Gamma^5n_{\beta i}^z
    +\Gamma^1n_{\beta i}^x+\Gamma^2n_{\beta i}^y
    \Bigr]c_i\,.
    \label{eq:H_U_magnon_FM}
\end{align}
As seen, the Hamiltonian is given by a replacement $\Gamma^5
\leftrightarrow \Gamma^{12}$, $\Gamma^1 \leftrightarrow \Gamma^{25}$,
$\Gamma^2 \leftrightarrow \Gamma^{51}$, and $\phi_{a0} \to \phi_{f0}$
in Eq.\,\eqref{eq:H_U_magnon_AFM}. Then the perturbative term $\delta
{\cal H}$ in the Hamiltonian ${\cal H}$ becomes,
\begin{align}
  \delta {\cal H} =
  \frac{U\phi_{f0}}{2}\Bigl[
    &-\Gamma^{12}(n^+_{\alpha i}n^-_{\alpha i}+
    n^+_{\beta i}n^-_{\beta i})
    \nonumber \\
    &+\Sigma^+n^-_{\alpha  i}+\Sigma^-n^+_{\alpha  i}
    +\Gamma^+n^-_{\beta  i}+\Gamma^-n^+_{\beta  i}
    \Bigl]\,.
\end{align}
Here we have omitted a term proportional to $\Gamma^5$ that is
ineffective for the FM order. The effective action is then obtained by
\begin{align}
  S_{\rm eff} \supset
  \frac{U\phi_{f0}^2}{2}\sum_{i\omega_n,\vb*{k}}\sum_{\xi=\alpha,\beta}
  \tilde{n}_{\xi}^+(k)\tilde{n}_{\xi}^-(-k)\tilde{\Gamma}^\xi_F(k)\,,
  \label{eq:Seff_F_mag}
\end{align}
and
\begin{align}
   \tilde{\chi}_F^{\alpha}(k)&=
  \frac{1}{2}\{\chi_F(k;\Sigma^+,\Sigma^-)
  +\chi_F(k;\Sigma^-,\Sigma^+)\}\,,\\
  \tilde{\chi}_F^{\beta}(k)&=
  \frac{1}{2}\{\chi_F(k;\Gamma^+,\Gamma^-)
  +  \chi_F(k;\Gamma^-,\Gamma^+)\}\,.
\end{align}
The first term of $\tilde{\Gamma}^\xi_F$ comes from the term
proportional to $\Gamma^{12}$.  It corresponds to the tadpole diagram
and it can be rewritten  by using Eq.\,\eqref{eq:phi_f0}. Thus the
structure is the same as $\tilde{\Gamma}_A^\xi$.  With the FM order,
the $\alpha$-type and $\beta$-type of magnon corresponds to $n_{\alpha
  i}^\pm$ and $n_{\beta i}^\pm$, respectively, which are shown in
Fig.\,\ref{fig:mag} (bottom).

From the expressions of the effective action regarding magnons, i.e.,
Eqs.\,\eqref{eq:Seff_A_mag} and \eqref{eq:Seff_F_mag}, the gap and
dispersion relation of the magnon is given by
\begin{align}
  \tilde{\Gamma}^{\xi}_M(i\omega_n=\omega+i\delta,\vb*{k}) = 0
  ~~~(\xi=\alpha,\beta,~~M=A,F)\,.
  \label{eq:gap_eq_magnon}
\end{align}

Now we show the numerical results of the dispersion relation of the
magnons. Fig.\,\ref{fig:omg_AFM-mag_AFM} shows a plot of the
dispersion relation regarding the $\alpha$-type magnon under the AFM
order.  We have found the dispersion relation $\omega(\vb*{k})$ is
minimized at $\vb*{k}=\vb*{0}$. On the other hand, there are
additional quasi-degenerate values of $\omega(\vb*{k})$ at
$\vb*{k}=(\pi,0,0),(0,\pi,0), (\pi,\pi,0)$. The degeneracy is about
1\%, compared to $\omega(\vb*{0})$ and the curvatures of the
wavenumbers are positive at the $\vb*{k}$s.  Therefore we expect four
different stable configurations of the magnon.  The excitation energy
is found to be around eV.\footnote{When we derive the effective action
for those configurations of magnon, we need to expand
$\tilde{\Gamma}^\alpha_A(\omega,\vb*{k})$ around each wavenumber,
i.e., $\vb*{k}=(\pi,0,0),(0,\pi,0), (0,\pi,\pi)$.}

\begin{figure*}[t]
  \begin{center}
    \includegraphics[scale=0.5]{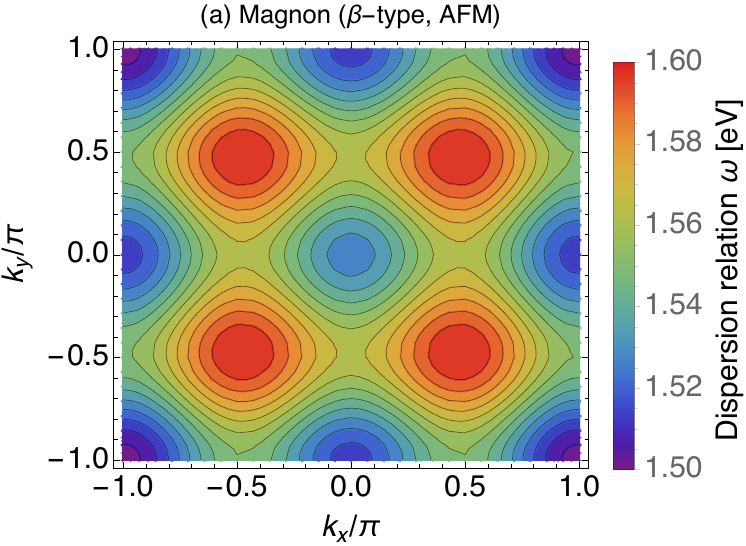}~~~~~
    \includegraphics[scale=0.5]{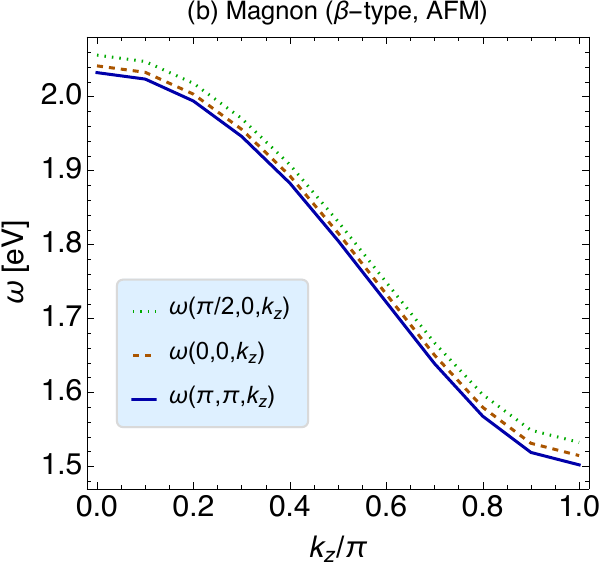}
  \end{center}
  \caption{Same as Fig.\,\ref{fig:omg_AFM-mag_AFM} but for $\beta$-type
    magnon under the AFM.  (a) Contour of
    $\omega(k_x,k_y,\pi)$. (b) $\omega(\vb*{k})$ as function of
    $k_z$ for various values of $k_x,k_y$.}
  \label{fig:omg_FM-mag_AFM}
\end{figure*}

Fig.\,\ref{fig:omg_FM-mag_AFM} shows the dispersion relation for the
$\beta$-type magnon under the AFM order. For the excitation similar
result to the $\alpha$-type  is obtained but with a different
configurations; we have found $\omega(\vb*{k})$ has the minimum at
$\vb*{k}=(\pi,\pi,\pi)$, and cross values are found at
$\vb*{k}=(0,0,\pi),(\pi,0,\pi),(0,\pi,\pi)$ with a degeneracy of
around 1\%. The energy scale turns out to be the same as the $\alpha$-type
one, i.e., $\order{\rm eV}$.

It is worth discussing the origin of the gap of the magnons. Let us
consider the $\alpha$-type magnon under the AFM at zero temperature.
Assuming that the expansion around $\omega=0$ is valid, we get the
stiffness and mass of the magnon as (see Appendix~\ref{app:L} for the
derivation)
\begin{align}
  J'_\alpha&
  =\int \frac{d^3\ell}{(2\pi)^3}\frac{2d^2-d_1^2-d_2^2}{8d^5}\,,
  \label{eq:J_alpha}\\
  J'_\alpha
  m_\alpha^{\prime \,2}&=\int \frac{d^3\ell}{(2\pi)^3}\frac{d_1^2+d_2^2}{2d^3}\,.
  \label{eq:Jm2_alpha}
\end{align}
It is clear that magnon is gapped at the zero temperature. From the
expression, we see that gap is typically $\order{\rm eV}$, which is
consistent with the numerical result shown in
Fig.\,\ref{fig:omg_AFM-mag_AFM}.  The typical scale is the same as the
AFM-type amplitude mode under the AFM. However, the parameter
dependence is different; in the $U\to \infty$ limit, $J'_\alpha
\propto {\rm eV}^3/U^3$ and $J'_\alpha m_\alpha^{\prime\, 2}\sim
A_2^2~{\rm eV}^2/U^3$ leads to $m'_\alpha\sim A_{2}$. Therefore, the
gap remains to be $\order{A_2}$ even if $U$ is much larger. We have
also confirmed this numerically.

Based on the above argument, we expect that the gap of the magnon
becomes zero when $d_1,d_2\to 0$. This can be confirmed without the
expansion with respect to $\omega$; considering the zero wavenumber
and taking the limit at the zero temperature,
Eq.\,\eqref{eq:gap_eq_magnon} with $M=A$ and $\xi=\alpha$ gives rise
to
\begin{align}
  1-\frac{U}{2N}\sum_{\vb*{\ell}}
  \frac{1}{d}\frac{1}{1-\omega^2/(4d^2)}=0\,.
\end{align}
Recalling Eq.\,\eqref{eq:phi_a0}, it is easy to find $\omega=0$ is the
solution to satisfy the gap equation. Therefore nonzero $d_1$ and
$d_2$ (or $A_2$) are the origin of the $\alpha$-type magnon gap under
the AFM.  This can be understood by considering the symmetry breaking
of the system.  When $d_1=d_2=0$, SO(3) symmetry in the
$\vec{\phi}_{Ai}-\vec{\phi}_{Bi}\equiv 2\vec{\phi}_{ai}$ space is
restored. Under the AFM order, the nonzero
$\vec{\phi}_{ai}=(0,0,\phi_{a0})$ breaks the SO(3) to
SO(2). Consequently two massless Nambu-Goldstone bosons appear, which
correspond to $n^\pm_{\alpha i}$. See Appendix~\ref{app:Symmetry} for
more detail.

Since we already have identified Nambu-Goldstone bosons, another
excitation, $\beta$-type magnon under the AFM order, should be massive
even if $d_1=d_2=0$. In fact, by expanding
$\tilde{\Gamma}_A^\beta(\omega,\vb*{0})$ around $\omega=0$, the
stiffness and mass are given by
\begin{align}
  J'_\beta&
  =\int \frac{d^3\ell}{(2\pi)^3}\frac{2d_5^2+d_1^2+d_2^2}{8d^5}\,,
  \label{eq:J_beta}\\
  J'_\beta
  m_\beta^{\prime\, 2}&=\int \frac{d^3\ell}{(2\pi)^3}
  \frac{2d_0^2-d_1^2-d_2^2}{2d^3}\,.
  \label{eq:Jm^2_beta}
\end{align}
Thus a finite mass (or gap) should appear even if $d_1,d_2\to 0$.

The results for $\alpha$-type and $\beta$-type magnons under
the FM state are similarly computed. It is found that qualitative
features of the $\alpha$-type and $\beta$-type ones are the same as
the $\beta$- and $\alpha$-types under the AFM order, respectively.
Typical energy scale of the excitations are a bit smaller but they are
$\order{\rm eV}$. See their plots in Appendix~\ref{app:mag_FM}.

To summarize this section, we have formulated the effective actions of
the amplitude modes and magnons under both the AFM and FM
orders. Each mode has two types; the amplitude mode has AFM- and
FM-types and the magnon has the $\alpha$- and $\beta$-types. From the
effective action, we have computed the dispersion relations for all
excitations. Among them, the AFM-type amplitude mode is identified as
the `axionic' quasi-particle. We have found that the mass of `axion'
coincides with the gap and it is derived more precisely compared to
the past works. The typical energy scale turns out to be eV and the
other excitations have the same scale.

\section{Implication to axion search}
\label{sec:axion_search}

As seen in Sec.\,\ref{sec:amp}, the `axionic' excitation corresponds
to the AFM-type amplitude mode (up to normalization of the field)
under the AFM, and it is shown its mass (or gap) $m_a$ is given by
$U\phi_{a0}$. On the other hand, the energy bands of the electrons of
the magnetic TIs are computed by the first-principles calculation.
For ${\rm Mn}_2{\rm Bi}_2{\rm Te}_5$, as an example,
Ref.\,\cite{Li:2020fvr} shows that the gap of the electrons under the
AFM order is $\order{10\,\mathchar`-\,10^2}$~meV.  With this
knowledge, we discuss the possibility of the  axion
detection by using the interaction of the `axion' with the
electromagnetic fields.

The `axion' field $\theta$ has an $\vb*{E}\vdot \vb*{B}$ coupling in
the Lagrangian~\cite{Li:2009tca}:
\begin{align}
  {\cal L}_{EB}= -\frac{\alpha}{4\pi}
   \theta F_{\mu\nu}\tilde{F}^{\mu\nu}
  =\frac{\alpha}{\pi}
 \theta 
  \vb*{E}\vdot \vb*{B} \,,
\end{align}
where $\alpha$ is the fine-structure constant and $F_{\mu\nu}$ is the
field strength of the electromagnetism.\footnote{$\mu,\nu,\rho,\sigma$
are the Lorentz indices that are 0, 1, 2, 3. }  $\tilde{F}^{\mu\nu}$
is the dual of $F_{\mu\nu}$, defined as
$\tilde{F}^{\mu\nu}=\frac{1}{2}\epsilon^{\mu\nu\rho\sigma}F_{\rho\sigma}$
($\epsilon^{\mu\nu\rho\sigma}$ is the Levi-Civita symbol with
$\epsilon^{0123}=+1$) and $\theta$ is related to the model parameters
of the 3D TIs as~\cite{Li:2009tca}
\begin{align}
  \theta = \frac{1}{4\pi} \int d^3k
  \frac{2d+d_4}{(d+d_4)^2d^3}
  \epsilon^{ijkl}d_i
  \partial_{k_x}d_j \partial_{k_y}d_k \partial_{k_z}d_l \,,
  \label{eq:theta}
\end{align}
with $i,j,k,l$ being 1, 2, 3, and 5.

\begin{figure*}[t]
  \begin{center}
    \includegraphics[scale=0.7]{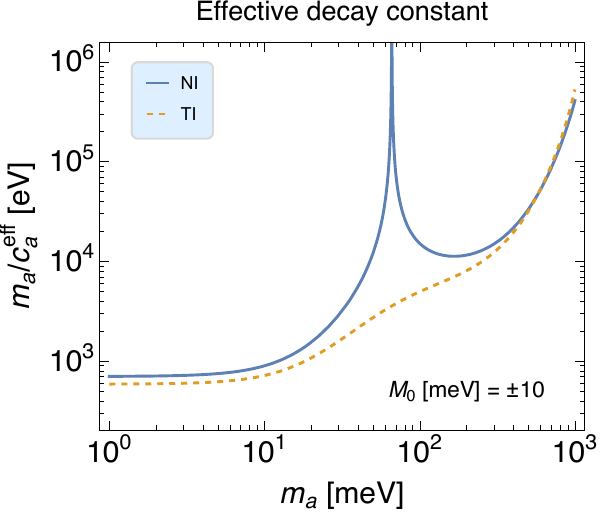}~~~~~
    \includegraphics[scale=0.7]{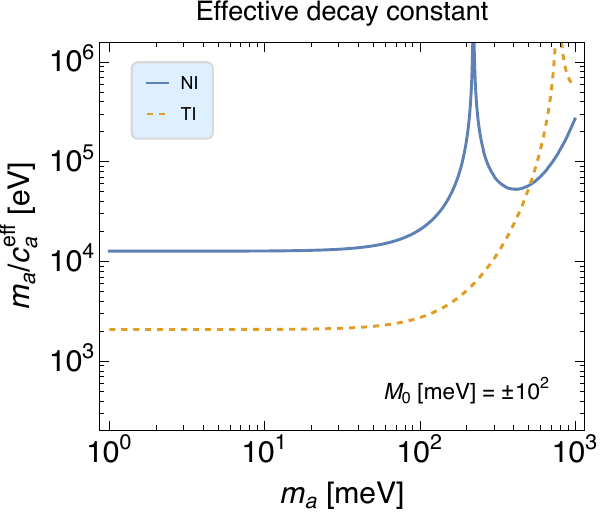}
  \end{center}
  \caption{Effective decay constant $m_a/c^{\rm eff}_a$ as function of
    gap $m_a$ of the amplitude mode. The lattice size is taken as
    $a=1\,{\rm \AA}$, $M_0\,[{\rm meV}]=\pm 10$ (left) and $\pm 10^2$
    (right) are taken, where positive (negative) $M_0$ corresponds to
    the NI (TI) phase. The other parameters ($A_1$, $A_2$, $B_1$ and
    $B_2$) are the same as Fig.\,\ref{fig:Veff}. }
  \label{fig:feff-amp}
\end{figure*}

Recalling that the effective action for the AFM-type amplitude mode
under the AFM is given in Eq.\,\eqref{eq:S_axion}, we define a
canonically normalized field $\delta \hat{\phi}_a$:
\begin{align}
  \frac{U}{2}\sqrt{J}\delta \phi_{a}\equiv
  \frac{1}{\sqrt{2}}\delta\hat{\phi}_a\,.
\end{align}
Therefore, using
\begin{align}
  \delta \theta &=\pdv{\theta}{d_5}\delta d_5
  \nonumber \\
  &=\pdv{\theta}{d_5}\frac{U}{2}\delta \phi_a
  \nonumber \\
  &=\pdv{\theta}{\ln d_5}\sqrt{\frac{2}{J}}\frac{1}{m_a}
  \delta \hat{\phi}_a\,,
\end{align}
the interaction term can be written in a form,
\begin{align}
  {\cal L}_{EB} \supset
  -\frac{\alpha }{4\pi}c_a^{\rm eff}
  \frac{\delta \hat{\phi}_a}{m_a}
  F_{\mu\nu}\tilde{F}^{\mu\nu}\,,
\end{align}
where 
\begin{align}
  c^{\rm eff}_a=\pdv{\theta}{\ln \phi_a}
 \sqrt{\frac{2}{J}}~\biggr|_{\phi_a=\phi_{a0}}\,.
\end{align}
Therefore, $m_a/c^{\rm eff}_a$ works as an effective``decay constant''
for axion.\footnote{Be aware that there is an additional factor of
$\alpha/(4\pi)$ in the $\delta \hat{\phi}_a F\tilde{F}$ coupling.} In
terms of a variable $g^{-1}=\pdv*{\theta}{m_5}$ (or
$\pdv*{\theta}{d_5}$ in our notation) in the
literature~\cite{Li:2009tca,Zhang_2020}, it is given by
\begin{align}
  g^2J=\frac{1}{2}\left(\frac{m_a}{c^{\rm eff}_a}\right)^2\,.
\end{align}
Or it is equivalent to  $f_Q$ used in Ref.\,\cite{Marsh:2018dlj}, i.e.,
\begin{align}
  f_Q=\frac{m_a}{c^{\rm eff}_a}\,.
\end{align}

We plot the effective decay constant of the amplitude mode as function
of $m_a$ in Fig.\,\ref{fig:feff-amp}. Here we have used the fact that
the AFM order parameter $\phi_{a0}$ becomes nonzero continuously as
$U$ increases shown in Sec.\,\ref{sec:Veff}. Therefore $m_a$ can take
any value from zero to $U$ if the value of $U$ is properly chosen.  In
the plot the results for the NI and TI phases are given by taking
$M_0=\pm 10,\,\pm 10^2$~meV for comparison. Here we note that the
enhancement point corresponds to a root of $c^{\rm eff}_a$. While the
results of the NI and TI are similar for $M_0=\pm 10$~meV, those with
$M_0=\pm 10^2$~meV are different. This quantitative difference comes
from the 3D TI model, which will be discussed below. To put it simply,
we found that the effective decay constant is
$\order{10^3\,\mathchar`-\,10^4}$\,eV for $m_a\lesssim 10^2$\,meV.

The result can be understood from a simple analytical estimation. Let
us focus on the case where $d_5\,(=m_a/2)=\order{1}$~meV that is
considered in Ref.\,\cite{Li:2009tca}.  To take an analytic approach,
we consider the Dirac model instead of the 3D TI model.  The Dirac
model is given by expanding the 3D TI model around $\vb*{k}=0$. See
Eq.\,\eqref{eq:d_a_Dirac} for the explicit form. It is empirically
known that physical properties of the excitation near its gap are
unchanged in either model.  Thus, the Dirac model can be a good tool
to understand the physical behavior analytically.  Since this scale is
much smaller than the energy scale of the electrons, $\theta$ in the
Dirac model is approximately given by~\cite{Ishiwata:2021qgd}
\begin{align}
  \theta \sim \frac{\pi}{2}[1-{\rm sgn}(M_0)]{\rm sgn}(d_5)
  +\frac{d_5}{M_0}\,,
\end{align}
which leads to 
\begin{align}
  g^{-1}\sim 1/M_0\,.
  \label{eq:g^-1}
\end{align}
In addition, $J$ is estimated as
\begin{align}
    J\simeq \int \frac{d^3\ell}{(2\pi)^3}\frac{1}{4d_0^3}
  \sim \order{10^{10}}\left(\frac{1\,{\rm \AA}}{a}\right)^3\,,
\end{align}
The estimation above is also checked numerically. Therefore, we get
\begin{align}
   \frac{m_a}{\sqrt{2}c^{\rm eff}_a}=\sqrt{g^2J}\sim
  \order{10^{5}}\times
  \left(\frac{1\,{\rm \AA}}{a}\right)^{3/2}|M_0|\,,
  \label{eq:g2J}
\end{align}
for $d_5\sim {\rm meV}$. This estimation is roughly consistent with
the results shown in Fig.\,\ref{fig:feff-amp}.  For a reference, we
have computed the effective decay constant with the Dirac model.  See
Appendix~\ref{app:mag_FM} for the plots of the results.

A careful reader may be curious about the result of the TI case with
$M_0=-100$~meV, which is relatively smaller than the rough estimation.
The suppression of the effective decay constant compared to the NI
case can be understood from Eq.\,\eqref{eq:theta}. In this expression,
only $d_4$ changes with the sign of $M_0$, and $d_4$ is slightly
suppressed when $M_0<0$ compared to $M_0>0$ case with $B_1$ and $B_2$
unchanged. As a result, $\theta$ tends to be slightly
enhanced. Therefore, since $\pdv*{\theta}{d_5}$ is also enhanced,
$m_a/c^{\rm eff}_a$ is suppressed. (On the other hand, for the case of
$|M_0|=10\,{\rm meV}$, the asymptotic value of the effective decay
constant does not depend on the sign of $M_0$ because the contribution
of $M_0$ itself in $d_4$ is smaller.)

As mentioned before, the enhancement of the effective decay constant
is due to the root of $c^{\rm eff}_a$. In general the root exists for
both the NI and TI cases and the location of the root depends on the
parameter. Therefore, although there are quantitative differences in
the behavior of effective coupling constants in NI and TI, depending
on the parameters, there is no qualitative difference in physical
properties.

Let us compare the result with the
literature. Ref.\,\cite{Marsh:2018dlj}, which proposes the axion
detection using `axion' in the magnetic TIs, claims that the effective
decay constant $f_Q$ is about $190$~eV, based on
Ref.\,\cite{Li:2009tca}.  This is smaller than our result. Let us see
this more closely.  In Ref.\,\cite{Li:2009tca}, they use a variable
$b^2\equiv \alpha^2B_0^2/2\pi^2\epsilon g^2J$,\footnote{We have
changed the expression in the literature to one in the natural unit.}
where $B_0=2$~T and $\epsilon=100$, and estimate $b$ as
$b=0.5$~meV. Using this relation, $f_Q=\sqrt{2g^2J}$ can be calculated
as $180$~eV, which is close to the value given in
Ref.\,\cite{Marsh:2018dlj}. Namely, the estimate of $b=0.5$~meV in
Ref.\,\cite{Li:2009tca} indicates $\sqrt{g^2J}\simeq 130$~eV. We find
this discrepancy comes from the input parameters, which will be
discussed below.

Some readers would be interested in the estimation of the axion mass
and the effective coupling by using the result of the first-principles
calculation. Applying the result of Ref.\,\cite{Li:2020fvr} to the
model parameters, i.e., $A_1=1.2\,{\rm eV}$, $A_2=2.6\,{\rm eV}$,
$B_1=-0.38\,{\rm eV}$, $B_2=-2.1\,{\rm eV}$, $M_0=-0.024\,{\rm eV}$,
we obtain $\phi_{a0}\simeq 1.5\times 10^{-3}$ and $m_a\simeq 20\,{\rm
  meV}$.  The effective decay constant is given by $m_a/c^{\rm
  eff}_a\simeq 3\times 10^2\,{\rm eV}$ for $a=1\,$\AA, which is
roughly consistent with he values referred in
Refs.\,\cite{Marsh:2018dlj,Schutte-Engel:2021bqm}.  This result is
qualitatively the same as $M_0=-10\,{\rm meV}$ case in
Fig.\,\ref{fig:feff-amp}. Additionally, the stability of the
excitation is checked. See Fig.\,\ref{fig:omg_AFM-amp_AFM_Li} for the
dispersion relation in Appendix~\ref{app:mag_FM}.  Here we point out
possible uncertainty that the determination of the parameter from the
first-principles calculation has. The first-principles calculation
provides the band structure of the material.  On the other hand, the
parameters of the model of the magnetic topological insulators can
only be determined by the fitting of the band structure.  For example,
the band gap determined by first-principles calculation gives the
value of $\sqrt{d_5^2+M_0^2}$, not the respective values of $d_5$ and
$M_0$. This difficulty comes from the magnetic order. If we consider
Bi$_2$Se$_3$, $d_5$ is zero due to the time reversal invariance from
the crystal structure. As a consequence, $M_0$ is determined by the
value of the band gap. In the present case, however, we consider the
state of the antiferromagnetic order and the above argument does not
apply. Since $d_5$ and $M_0$ play an important role in determining the
axion mass and the effective coupling constant, this indeterminacy
will likely strongly affect them.

To summarize this section, the effective decay constant ranges as
$\order{10^2\,\mathchar`-\,10^4}$\,eV, depending on the parameters. In
addition, we have shown that the effective decay constant is
qualitatively topology-independent.  This fact will encourage
selecting suitable material for axion search from a broad perspective.

Finally, we point out a possible interaction of the magnon with the
electromagnetic fields.  For instance, since $\alpha$-type magnon
appears as $\Gamma^5$ term, we have
\begin{align}
  \delta \theta &=\pdv{\theta}{d_5}\delta d_5
  \nonumber \\
  &=\pdv{\theta}{d_5}\left(-\frac{U}{4}\phi_{a0}n_\alpha^+n_\alpha^-\right)
  \nonumber \\
  &=-\pdv{\theta}{\ln d_5}\frac{2}{J_\alpha}\frac{1}{m_a^2}
  \hat{n}_\alpha^+\hat{n}_\alpha^-\,,
\end{align}
where we have defined a canonically normalized field for the
$\alpha$-type magnon under the AFM order as
\begin{align}
  \frac{U\phi_{a0}}{2}\sqrt{J_\alpha}n_{\alpha}^\pm\equiv
  \hat{n}^\pm_\alpha\,.
\end{align}
Then the interaction with the electromagnetic field is
\begin{align}
  {\cal L}_{EB} \supset
  -\frac{\alpha }{4\pi}c_\alpha^{\rm eff}
  \frac{\hat{n}^+_\alpha\hat{n}^-_\alpha}{m^2_a}
  F_{\mu\nu}\tilde{F}^{\mu\nu}\,,
\end{align}
where
\begin{align}
  c^{\rm eff}_\alpha=\pdv{\theta}{\ln \phi_a}
 \frac{2}{J_\alpha}~\biggr|_{\phi_a=\phi_{a0}}\,.
\end{align}
The interaction indicates that two magnons are excited under the
electromagnetic fields, which can be another interesting signal for
the axion detection. However, we have found that the magnon tends to
dissipate when its mass is below $\order{{\rm eV}}$. This is true when
we use the parameters given by Ref.\,\cite{Li:2020fvr}. Therefore, it
would be challenging to detect such excitations.  Instead, there may
be a possibility of finding a relic of the dissipation of the two
magnons, which is induced by the axion.  We leave it for future
investigation.

\section{Conclusion}
\label{sec:conclusion}

We have investigated possible excitations in the effective model of 3D
magnetic topological insulators and discussed their impact on the
axion detection. The model consists of the 3D effective model of TIs
and the Hubbard term. In the current study we focus on the zero
temperature case. Computing the effective potential from the grand
potential, the AFM and FM orders are found. Both orders are triggered
by a large value of $U$, which is a dimensionful coefficient of the
Hubbard term. It turns out that the AFM state is lower energy than the
FM, i.e., the AFM is the global minimum of the system.  The order
parameter $\phi_{a0}$ of the AFM appears continuously from zero,
meanwhile that of the FM state turns out to emerge discontinuously.

Under the magnetic orders, we have derived the effective action for
possible magnetic excitations. The magnetic excitations are classified
by amplitude mode and magnon. Furthermore, the amplitude mode and
magnon have two types: AFM/FM-types and $\alpha$/$\beta$-types.  We
consider those four excitations under both the AFM and FM states, that
are in total eight magnetic excitations. We found that the effective
actions for all excitations are given by the inverse propagator that
is composed of the dynamical susceptibility.  The formulation provides
not only the dispersion relation and the correspondence between the
gap and mass of the excitations but also a criterion for the stability
and validity of the effective description of the excitations.

With the formalism, we have found the AFM-type amplitude mode under
the AFM is stable and the other amplitude modes tend to dissipate.
The gap (or mass) of the AFM-type amplitude mode under the AFM is
given by $m_a=U\phi_{a0}$. Namely, the gap is typically $U\,(\sim$eV) for a
saturated magnetization $\phi_{a0}\sim 1$. On the other hand, it can be
suppressed when $\phi_{a0}\ll 1$.

The four magnons turn out to be stable and their typical mass scale is
$\order{\rm eV}$. For stable magnons, we have found there are several
quasi-degenerate configurations. Taking the $\alpha$-type magnon under
the AFM as an example, the states with wavenumber $(\pi,0,0)$,
$(0,\pi,0)$, and $(\pi,\pi,0)$ are found to be stable in addition to
one with $(0,0,0)$. On the other hand, we discovered the magnon tends
to dissipate for $\phi_{a0}\lesssim 1$.

We are especially interested in the AFM-type amplitude modes because
they relate to `axionic' quasi-particle and axion detection. First of
all, we have identified `axion' as the AFM-type amplitude mode under
the AFM.  Besides, we have determined its mass and the effective
coupling to the electromagnetic fields more accurately than the past
works.  Since the mass ranges from zero to $U$, it is possible to
consider a situation where the mass is less than
$\order{10^2}$\,meV. In the scale, the effective decay constant of
‘axion,’ determined by the effective coupling and the mass, turns out
to be $\order{10^2\,\mathchar`-\,10^4}$\,eV, depending on the
parameters. This value is roughly the same as, or up to two orders of
magnitude larger than, the previous estimate, which has a significant
impact on the proposal of the axion detection using ‘axion’ in the
magnetic TIs. We also point out that the nature of `axion' is
insensitive to the topology of the magnetic insulators, i.e.,
topological phase or normal phase.

As shown, the topological nature of the materials is unnecessary to
utilize `axion' for the search.  The magnon under the AFM state is
another possibility to access to axion. Therefore, it is worth
pursuing materials from broader candidates by studying their magnetic
states and collective excitations that are suitable for the axion
search. On the theoretical side, extensions of the model and
simulations by the first-principles calculation would be the next step
to further investigation of the magnetic excitations.  We will leave
it for future study.

\begin{acknowledgments}
  We are grateful to Makoto Naka for valuable discussions in the early
  stage of this project. We also thank Fumiyuki Ishii, Kaiki Shibata,
  and Naoya Yamaguchi for useful discussion.  This work was supported
  by JSPS KAKENHI Grant Numbers JP18H05542, JP20H01894, and JSPS
  Core-to-Core Program Grant No. JPJSCCA20200002 (KI), and JST CREST,
  Grant No. JPMJCR18T2 (KN).
\end{acknowledgments}

\appendix

\section{Notation}
\label{app:notation}

We summarize the Gamma matrices and the notation for the wavefunction
and Green's function we use in this paper.

\subsection{Gamma matrices}
\label{app:Gammas}

In this study we mainly use the Gamma matrices in the so-called sublattice
basis.  In the basis the Gamma matrices $\Gamma^a$ ($a=1,\cdots, 5$)
are defined as
\begin{align}
  \Gamma^{1}=\mqty(\sigma^1 & 0 \\ 0 & -\sigma^1)\,,~
  &\Gamma^{2}=\mqty(\sigma^2 & 0 \\ 0 & -\sigma^2)\,,~
  \Gamma^{3}=\mqty(0 & -i\bf{1} \\ i\bf{1} &0)\,,~
  \nonumber \\
  \Gamma^{4}=&\mqty(0 & -\bf{1} \\ -\bf{1} &0)\,,~
  \Gamma^{5}=\mqty(\sigma^3 &0 \\ 0 &-\sigma^3)\,.
\end{align}
$\Gamma^5$ can also be written by
$\Gamma^5=-\Gamma^1\Gamma^2\Gamma^3\Gamma^4$.  In addition, we define
\begin{align}
  \Gamma^{ab}=\frac{1}{2i}\,[\Gamma^a,\Gamma^b]\,.
  \label{eq:Gamma^ab}
\end{align}
The explicit form of $\Gamma^{25}$, $\Gamma^{51}$, and $\Gamma^{12}$
are
\begin{align}
  \Gamma^{25}=\mqty(\sigma^1 &0 \\ 0 &\sigma^1)\,,~
  \Gamma^{51}=\mqty(\sigma^2 &0 \\ 0 &\sigma^2)\,,~
  \Gamma^{12}=\mqty(\sigma^3 &0 \\ 0 &\sigma^3)\,.
  \label{eq:Sigma_a}
\end{align}
The operators $\Gamma^a$ ($a=1,2,5$) and $\Gamma^{ab}$ ($ab=25,51,12$)
represent the AFM and FM order parameters of $(x,y,z)$ directions,
respectively. Especially $\Gamma^{ab}$ ($ab=25,51,12$) correspond to
the spin operator in the Dirac Gamma matrices in particle physics.

Recalling the Dirac Gamma matrices, one may wonder that $\Gamma^{ab}$
($ab=23,31,12$) should be used instead of
Eq.\,\eqref{eq:Sigma_a}. This confusion is due to different
representations of the Gamma matrices.  Namely, $\Gamma^a$
($a=1,2,3,5$) can be replaced each other by a unitary transformation.
For instance, a unitary transformation $\Gamma ^a \to
\Gamma^{a\,\prime} = U \Gamma^a U^\dagger$ by a unitary matrix
\begin{align}
  U = \frac{1}{\sqrt{2}}\mqty(1 & -i \sigma^1 \\ -i\sigma^1 & 1)\,,
\end{align}
which is $U_3$ in Ref.\,\cite{Chigusa:2021mci}, gives $\Gamma^3\to
\Gamma^5$, $\Gamma^5 \to -\Gamma^3$, and the others remain.  This
indicates that we can construct a representation of the Lorentz group
from three matrices out of four ones. In the sublattice basis, we can
use $\Gamma^1$, $\Gamma^2$, and $\Gamma^5$. Let us see this explicitly
below.

To begin with, we define $J_i$ ($i=1,2,3$) and $K_i$ ($i=1,2,3$) as
\begin{align}
  J_i=\frac{1}{4}\epsilon_{iab}\Gamma^{ab}\,, ~
  K_i=\frac{i}{4} \Gamma^{4i}\,,
  \label{eq:J_iK_i}
\end{align}
where $\epsilon_{iab}$ is the Levi-Civita symbol and indices $a$ and
$b$ take 1, 2 or 5, such as $J_1=\Gamma^{25}/2,J_2=\Gamma^{51}/2$
etc. Then they obey the following commutation relation:
\begin{align}
  [J_i,J_j]=i\epsilon_{ijk}J_k\,,~
  [K_i,K_j]=-i\epsilon_{ijk}J_k\,,~
  [J_i,K_j]=i\epsilon_{ijk}K_k\,,
\end{align}
which shows that $J_i$ and $K_i$ are the generators of the Lorentz
transformation. Especially $J_i$ are the operators for the rotational
transformations.  Let us focus on $J_1$ as an example. An operator for a
finite $\theta$ rotation along $i=1$, i.e., $x$ axis, is given by
\begin{align}
  S=e^{i\frac{\theta}{2}\Gamma^{25}}
  =\cos \frac{\theta}{2}+i\Gamma^{25}\sin \frac{\theta}{2}\,,
\end{align}
Then we obtain
\begin{align}
  S^{-1}\Gamma^1 S &= \Gamma^1\,,
  \\
  S^{-1}\Gamma^2 S &= \Gamma^2 \cos\theta +\Gamma^5 \sin \theta\,,
  \\
  S^{-1}\Gamma^5 S &= -\Gamma^2 \sin\theta +\Gamma^5 \cos \theta\,.
\end{align}
Therefore, it is confirmed that $J_i$ are the generators for rotation
in three dimensional space with axes $i=1,2,5$. Since the generators
of the rotational transformation are the spin operators up to a
factor, $\Gamma^{25}$, $\Gamma^{51}$, and $\Gamma^{12}$ are introduced
for the FM order. Accordingly, $\Gamma^{1}$, $\Gamma^2$,
and $\Gamma^5$ are used for the AFM order.

We note that the above calculation of the rotational transformation of
the Gamma matrices is just to show an example that
Eq.\,\eqref{eq:J_iK_i} are the generators of the Lorentz
transformation. This has nothing to do with the discrete symmetry of
the crystal structure. The 3D TI model has the time-reversal symmetry,
space inversion symmetry, and three-fold rotation symmetry along the
$z$ axis~\cite{Zhang:2009zzf,Li:2009tca,Liu_2010}, except for
$\Gamma^5$ term.  Each Gamma matrix is the representation of the
symmetries~\cite{Liu_2010} and the unitary transformation of the Gamma
matrices does not change the physical interpretation of each
$\Gamma^a$.

\subsection{Wavefunction and Green's function}
\label{app:Fourier_ex}

In a finite temperature the wavefunction $\Phi_i$ and the Green's function
$G$ is Fourier expanded as
\begin{align}
   \Phi_{i}(\tau)&=\frac{1}{\sqrt{\beta N}}\sum_{i\omega_n,\vb*{k}}
   \tilde{\Phi}(i\omega_n,\vb*{k})e^{-i\omega_n\tau+i\vb*{k}\vdot \vb*{x}_i}\,,
   \\
   G(x_i-x_j)&=\frac{1}{\beta N}\sum_{i\omega_n,\vb*{k}}
   \tilde{G}(i\omega_n,\vb*{k})
   e^{-i\omega_n(\tau_i-\tau_j)+i\vb*{k}\vdot(\vb*{x}_i-\vb*{x}_j)}\,,
\end{align}
where $\vb*{x}_i$ shows the location of the cite $i$ and
$x_i=(\tau_i,\vb*{x}_i)$. $\Phi_i$ includes $c_i$, $\delta\phi_{ai}$,
$\delta\phi_{fi}$, and $n^\pm_{\alpha i}$, $n^\pm_{\beta
  i}$. Regarding the Green's function, we change the definition of $G$
according to the magnetic order of the system. For instance, if there
is no magnetic order, then we choose $G^{-1}=-\partial_\tau-{\cal
  H}^{\rm TI}$, leading to $\tilde{G}=(-i\omega_n+d_0)^{-1}$.

\section{Dynamical susceptibility}
\label{app:chi}

In this section, we give the expression for the dynamical
susceptibility used in the numerical calculation. We have introduced
eight kinds of susceptibilities and they can be written in a generic
form as
\begin{align}
  \chi_M(k;O_1,O_2)
   =-\frac{1}{\beta N}\sum_{i\omega_\ell,\vb*{\ell}}
  {\rm Tr}[\tilde{G}_M(\ell)O_1\tilde{G}_M(\ell+k)O_2]\,,
\end{align}
where $O_1$ and $O_2$ stand for the Gamma matrices, such as
$\Gamma^a$, $\Gamma^\pm$, or $\Sigma^\pm$. Recalling that
$\tilde{G}_M(\ell)=(-i\omega_{\ell}+{\cal H}_{\vb*{\ell}})^{-1}$,
$\tilde{G}_M(\ell)$ can be diagonalized by a unitary matrix as 
\begin{align}
  U_{\vb*{\ell}}^\dagger  \tilde{G}_M(\ell) U_{\vb*{\ell}}
  =\tilde{G}_M(\ell)_{\rm diag}\,,
\end{align}
where $\tilde{G}_M(\ell)_{\rm diag}$ is a diagonal matrix and
$\tilde{G}^{-1}_M(\ell)={\rm
  diag}(-i\omega_\ell+E_{1\vb*{\ell}},-i\omega_\ell+E_{2\vb*{\ell}},
-i\omega_\ell+E_{3\vb*{\ell}},-i\omega_\ell+E_{4\vb*{\ell}})$. Similarly,
$\tilde{G}_M(\ell+k)$ is diagonalized as
\begin{align}
  U_{\vb*{\ell}+\vb*{k}}^\dagger  \tilde{G}_M(\ell+k) U_{\vb*{\ell}+\vb*{k}}
  =\tilde{G}_M(\ell+k)_{\rm diag}\,.
\end{align}
Then the trace part is
\begin{align}
  {\rm Tr}[\tilde{G}_M(\ell)O_1\tilde{G}_M(\ell+k)O_2]
  &={\rm Tr}[\tilde{G}_M(\ell)_{\rm diag}O'_1
    \tilde{G}_M(\ell+k)_{\rm diag}O'_2]
  \nonumber \\
  &=\sum_{i,j}
  \frac{1}{i\omega_n-E_{i\vb*{k}}}
  \frac{1}{i\omega_n+i\omega_\ell-E_{j\vb*{\ell}+\vb*{k}}}
  (O'_1)_{ij}(O'_2)_{ji}
\end{align}
where
\begin{align}
  O'_1 &= U_{\vb*{\ell}}^\dagger O_1 U_{\vb*{\ell}+\vb*{k}}\,,
  \\
  O'_2 &= U_{\vb*{\ell}+\vb*{k}}^\dagger O_2 U_{\vb*{\ell}}\,.
\end{align}
Summing over $i\omega_\ell$, we finally get
\begin{align}
  \chi_M(i\omega_n,\vb*{k};O_1,O_2)
   =-\sum_{\vb*{\ell}}
  \sum_{i,j}
  \frac{(O'_1)_{ij}(O'_2)_{ji}}
       {i\omega_n+E_{j\vb*{\ell}+\vb*{k}}-E_{i\vb*{k}}}
       (n_F(E_{i\vb*{\ell}})-n_F(E_{j\vb*{\ell}+\vb*{k}}))\,.
       \label{eq:chi_sum}
\end{align}

\section{Dynamical susceptibility under AFM at zero temperature}
\label{app:chi_AFM}

With the AFM order at zero temperature, the susceptibility can be
derived analytically. In the limit $\beta^{-1}\sum_{i\omega_\ell}$ is
replaced by $\int d\ell_E^0/(2\pi)$ where $\omega_\ell =\ell^0_E$. We
list the results below:
\begin{align}
  \tilde{\chi}_A^a(i\omega_n=\omega,\vb*{k})
  &=
  \frac{2}{N}\sum_{\vb*{\ell}}
  \frac{df+\sum_{a=1}^5d_af_a-2d_5^2}{P}\,,
  \label{eq:chi_A^a}
  \\
  \tilde{\chi}_A^{f}(i\omega_n=\omega,\vb*{k})
  &=\frac{2}{N}\sum_{\vb*{\ell}}
  \frac{df-\sum_{a=1}^5d_af_a+2(d_1f_1+d_2f_2)}{P}\,,
  \\
  \tilde{\chi}_A^{\alpha}(i\omega_n=\omega,\vb*{k})
  &=
  \frac{2}{N}\sum_{\vb*{\ell}}\frac{df+\sum_{a=1}^5d_af_a
    -(d_1f_1+d_2f_2)}{P}\,,
  \\
  \tilde{\chi}_A^{\beta}(i\omega_n=\omega,\vb*{k})
  &=
  \frac{2}{N}\sum_{\vb*{\ell}}
  \frac{df-\sum_{a=1}^5d_af_a+(2d_5^2+d_1f_1+d_2f_2)}{P}\,,
\end{align}
where
\begin{align}
  P=df(d+f)\{1-\omega^2/(d+f)^2\}\,.
\end{align}
$d_a$ and $f_a$ $(a=1$\,--\,$4)$ depend on the wavenumber as
$d_a=d_a(\vb*{\ell})$ and $f_a=d_a(\vb*{\ell}+\vb*{k})$ and
$f_5=d_5$. $f$ is defined by $f=(\sum_{a=1}^5f_af_a)^{1/2}$.

\section{How to derive the effective action of the excitation
  at zero temperature}
\label{app:L}

In this section we derive the effective action for the excitation in
the continuum and zero temperature limit.  Using the dynamical
susceptibility, the effective (Euclidean) action for the excitation
has a form (see Eqs.\,\eqref{eq:Samp}, \eqref{eq:Seff_A_mag} and
\eqref{eq:Seff_F_mag}),
\begin{align}
  S_E = \frac{U\phi_0}{2} \sum_{i\omega,\vb*{k}}
  \tilde{\eta}^*(k)\tilde{\eta}(-k)
  \tilde{\Gamma}(k)\,,
\end{align}
where $\eta$ is the excitation ($\tilde{\eta}$ is the Fourier
coefficient) and
\begin{align}
  \tilde{\Gamma}(k)\equiv
  1-\frac{U}{4}\tilde{\chi}_M(i\omega_n,\vb*{k})\,.
\end{align}
$\phi_0$ is unity, $\phi^2_{a0}$, or $\phi^2_{f0}$, depending on the
excitation. Since $\tilde{\Gamma}$ is the inverse of the two-point
Green's function, the effective action can be derived by expanding it
at the mass $m$,
\begin{align}
  \tilde{\Gamma}(i\omega_n,\vb*{k})
  &=\pdv{\tilde{\Gamma}}{(i\omega_n)^2}
  \Bigl|_{i\omega_n=m,\vb*{k}=\vb*{0}}((i\omega_n)^2-m^2)
  \nonumber \\
  &+\sum_i
  \pdv{\tilde{\Gamma}}{k_i^2}\Bigl|_{i\omega_n=m,\vb*{k}=\vb*{0}}k_i^2
  +\cdots\,.
\end{align}
Here we have used the definition of the mass:
\begin{align}
  \tilde{\Gamma}(i\omega_n=m,\vb*{0})=0\,.
\end{align}
In the continuum coordinate space it can be written as
\begin{align}
  S_E &= \frac{U\phi_0}{2} \int^\beta d\tau \sum_i
  \eta^*\frac{U}{2}a^3 J (-\partial_\tau^2+v_i^2k_i^2+m^2)\eta
  \nonumber \\
  &~\to~~ \frac{U\phi_0}{2} \int^\beta d\tau \int d^3x
  \eta^*\frac{U}{2}J (-\partial_\tau^2+v_i^2k_i^2+m^2)\eta\,,
\end{align}
where we have retrieved the lattice size $a$ to make the dimension of
the action intact.  $J$ and $v_i$ are the stiffness and velocity
defined as
\begin{align}
  &\frac{U}{2}a^3J=-\pdv{\tilde{\Gamma}}{(i\omega_n)^2}
  \Bigl|_{i\omega_n=m,\vb*{k}=\vb*{0}}\,,
  \\
  &\frac{U}{2}a^3Jv_i^2=\pdv{\tilde{\Gamma}}{k_i^2}
  \Bigl|_{i\omega_n=m,\vb*{k}=\vb*{0}}\,.
\end{align}
Using the real time ($t=-i\tau$), we get the (Minkowski) action as
\begin{align}
  iS_E ~~\to~~ S= \left(\frac{U}{2}\right)^2\phi_0 J\int d^4x 
  \eta^* (-\partial_t^2+v_i^2\partial_i^2-m^2)\eta\,.
\end{align}

\begin{figure}[t]
  \begin{center}
    \includegraphics[scale=0.7]{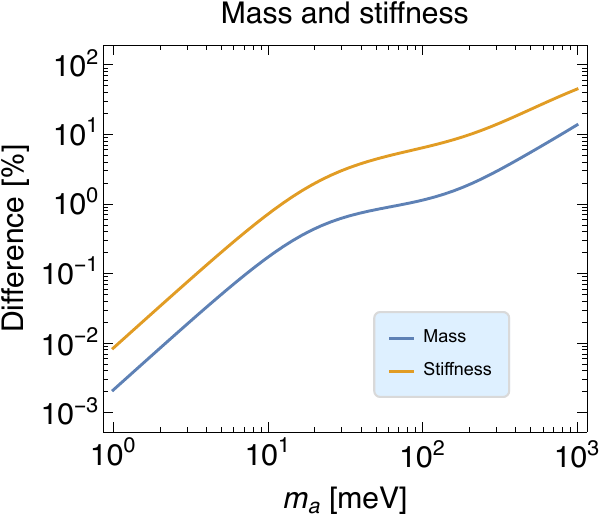}
  \end{center}
  \caption{Comparison of the mass and stiffness of axionic
    excitation. $r_m=-(m_a-m_a^{\rm old})/m_a$ and $r_J=(J-J^{\rm
      old})/J$ are plotted as function of $m_a$.  The other parameters
    are the same as Fig.\,\ref{fig:Veff}.}
  \label{fig:rs}
\end{figure}

Let us calculate the stiffness for the AFM-type amplitude mode in the
AFM order. In this case, the mass is given by $U\phi_0$. Then
\begin{align}
  \frac{U}{2}a^3J
 & =\frac{U}{2N}\sum_{\vb*{\ell}} \frac{d_0^2}{4d^5}
  \frac{1}{(1-d_5^2/d^2)^2}
  \nonumber \\
  \to ~~ &J =\int \frac{d^3\ell}{(2\pi)^3}\frac{1}{4dd_0^2}\,.
\end{align}
In the previous works~\cite{Li:2009tca,Ishiwata:2021qgd},
$\tilde{\Gamma}$ is expanded around $i\omega_n=0$. Namely,
\begin{align}
  \tilde{\Gamma}^a_A(i\omega_n=\omega,\vb*{0})
  =
  \frac{U}{2N}\sum_{\vb*{\ell}}\left[
    \frac{d^2_5}{d^3}-\frac{d_0^2}{4d^5}\omega^2 +\order{\omega^4}
    \right]\,.
  \label{eq:expansion_old}
\end{align}
As a result, we get
\begin{align}
  J^{\rm old}&=\int \frac{d^3\ell}{(2\pi)^3}\frac{d^2_0}{4d^5}\,,
  \label{eq:app:J_old}
  \\
  J^{\rm old} (m_a^{\rm old})^2
  &=\int \frac{d^3\ell}{(2\pi)^3}\frac{d_5^2}{d^3}\,.
  \label{eq:app:Jma_old}
\end{align}
It is clear that the expansion around $\omega=0$ in
Eq.\,\eqref{eq:expansion_old} is guaranteed if $m_a^{\rm old}$ is
smaller than $2d$. If this condition is satisfied, then the expansion
is valid and we expect $m_a^{\rm old}\simeq m_a$. The discrepancy
between the old results and the new ones is shown in
Fig.\,\ref{fig:rs}. We found that they agree at around
$(10^{-3}\,\mathchar`-\,10)$\% for $m_a\sim$\,(meV - eV). Therefore
the old results are a good approximation for $m_a\lesssim \order{\rm
  eV}$.

Similarly, the stiffness for the $\alpha$-type magnon with the AFM is
\begin{align}
  J_\alpha = \int \frac{d^3\ell}{(2\pi)^3}
  \frac{2d^2-d_1^2-d_2^2}{8d^3}\frac{1}{(1-m_\alpha^2/(4d^2))^2}\,,
\end{align}
where the mass $m_\alpha$ is given by
$\tilde{\Gamma}_A^\alpha(i\omega_n=m_\alpha,\vb*{0})=0$.  On the other
hand, if we expand $\tilde{\Gamma}^\alpha_A$ around $i\omega_n=0$, i.e.,
\begin{align}
  \tilde{\Gamma}^\alpha_A(i\omega_n=\omega,\vb*{0})
  =\frac{U}{2N}\sum_{\vb*{\ell}}\left[
    \frac{d^2_1+d^2_2}{2d^3}
    -\frac{2d^2-d_1^2-d_2^2}{8d^5}\omega^2 +\order{\omega^4}
    \right]\,,
\end{align}
we obtain
\begin{align}
  J'_\alpha&
  =\int \frac{d^3\ell}{(2\pi)^3}\frac{2d^2-d_1^2-d_2^2}{8d^5}\,,
  \label{eq:app:J_alpha}\\
  J'_\alpha
  m_\alpha^{\prime 2}&=\int \frac{d^3\ell}{(2\pi)^3}\frac{d_1^2+d_2^2}{2d^3}\,.
  \label{eq:app:Jm2_alpha}
\end{align}

\section{Symmetry of the Hamiltonian}
\label{app:Symmetry}

To understand the appearance of the gapless modes, we use
\begin{align}
  \vec{\phi}_{ai}&=(\vec{\phi}_{Ai}-\vec{\phi}_{Bi})/2\,,
  \\
  \vec{\phi}_{fi}&=(\vec{\phi}_{Ai}+\vec{\phi}_{Bi})/2\,,
\end{align}
in this section.  The Hubbard term after the Stratonovich-Hubbard
transformation gives,
\begin{align}
  H_U \to
  \frac{U}{2}\sum_{i}(\vec{\phi}_{ai}^{\,2}+\vec{\phi}_{fi}^{\,2})+
  \frac{U}{2}\sum_{i}c^\dagger_i
  \Bigl[
    \vec{\phi}_{ai}\vdot \vec{\Gamma}
    +\vec{\phi}_{fi}\vdot \vec{\Sigma}
    \Bigr]c_i\,,
\end{align}
where we have introduced
$\vec{\Gamma}=(\Gamma^1,\Gamma^2,\Gamma^5)$ and
$\vec{\Sigma}=(\Gamma^{25},\Gamma^{51},\Gamma^{12})$.  It can
be seen that the rotational invariance of the Hubbard term is intact
in the parameter space of magnetic orders, $\vec{\phi}_{ai}$ and
$\vec{\phi}_{fi}$, which corresponds to $(\Gamma^1,\Gamma^2,\Gamma^5)$
space.  See also Appendix~\ref{app:Gammas}. This symmetry is broken by
${\cal H}^{\rm TI}$, i.e., $\Gamma^1$ and $\Gamma^2$ terms that couple
to the wavenumbers. To see the structure of the symmetry, we take
$d_1=d_2=0$ hereafter. In addition we consider a uniform magnetic
configuration to study the magnons of the zero mode.

The Hamiltonian ${\cal H}$ is given by
\begin{align}
  {\cal H}
  &={\cal H}^{\rm MTI}+\delta {\cal H} \nonumber
  \\
  &=\sum_{a=3}^4
  d_a\Gamma^a+\frac{U}{2}(\vec{\phi}_{a}\vdot \vec{\Gamma}
    +\vec{\phi}_{f}\vdot \vec{\Sigma})\,.
\end{align}
and the energy eigenvalues are 
\begin{align}
  \pm \sqrt{d_s^2+\vec{\phi}_a^2+\vec{\phi}_f^2
    +2\sqrt{\vec{\phi}_a\vdot \vec{\phi}_f
      +d_s^2(\vec{\phi}_a^2+\vec{\phi}_f^2)}}\,,
  \\
  \pm \sqrt{d_s^2+\vec{\phi}_a^2+\vec{\phi}_f^2
    -2\sqrt{\vec{\phi}_a\vdot \vec{\phi}_f
      +d_s^2(\vec{\phi}_a^2+\vec{\phi}_f^2)}}\,.
\end{align}
The expression clearly shows that the system has SO(3) symmetry.
Under the AFM state, with $(\phi_{az},\phi_{fz})=(\phi_{a0},0)$, you
can see that SO(3) is broken to SO(2). Therefore,  massless
Nambu-Goldstone bosons are expected, which corresponds to the
$\alpha$-type magnon. The symmetry breaking by the FM order is the
same. In this case, the $\alpha$-type magnon is the Nambu-Goldstone
bosons.

\begin{figure*}[t]
  \begin{center}
    \includegraphics[scale=0.5]{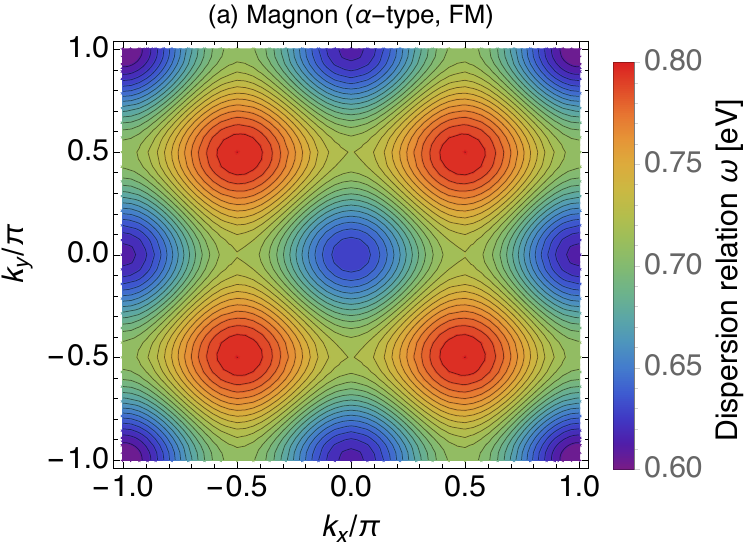}~~~~~
    \includegraphics[scale=0.5]{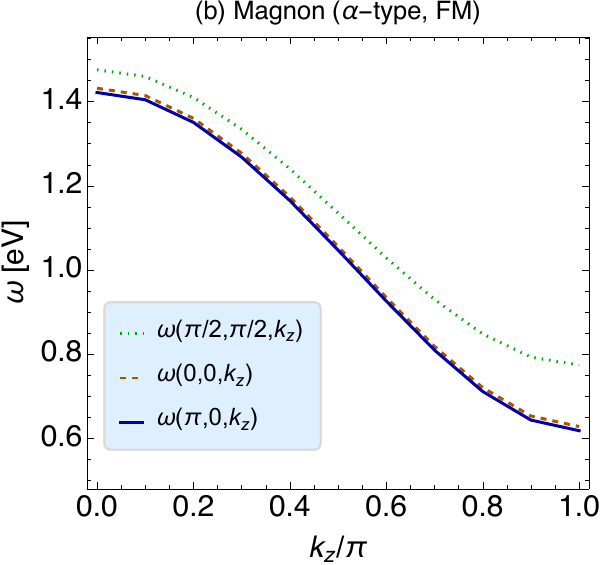}
  \end{center}
  \caption{Dispersion relation for $\alpha$-type magnon under the FM.
    (a) Contour of $\omega(k_x,k_y,\pi)$. (b) $\omega(\vb*{k})$ as
    function of $k_z$ for various values of $k_x,k_y$.}
  \label{fig:omg_FM-mag_FM}
\end{figure*}

\begin{figure*}[t]
  \begin{center}
    \includegraphics[scale=0.5]{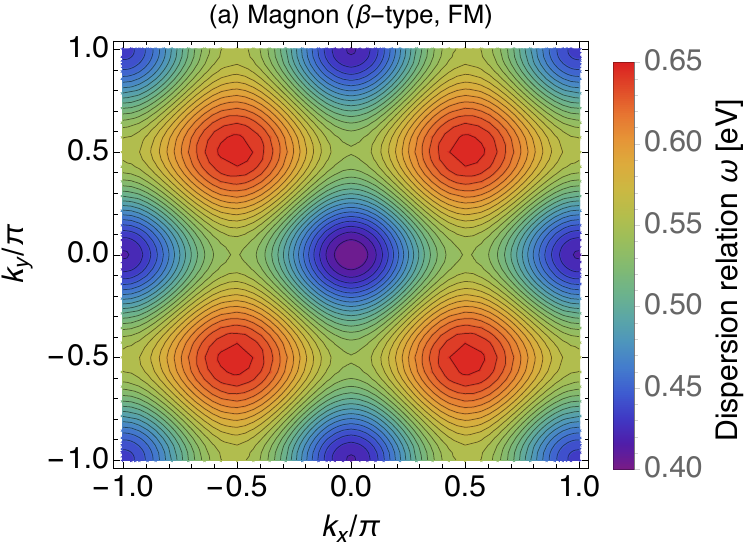}~~~~~
    \includegraphics[scale=0.5]{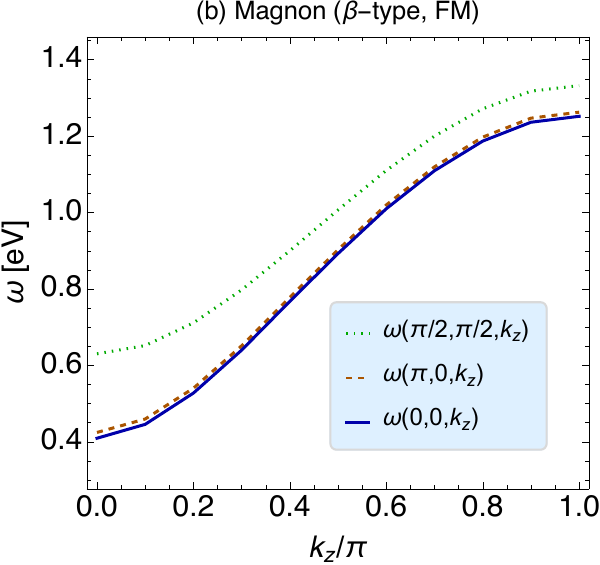}
  \end{center}
  \caption{Same as Fig.\,\ref{fig:omg_FM-mag_FM} but for $\beta$-type
    magnon under the FM.  (a) Contour of $\omega(k_x,k_y,0)$. (b)
    $\omega(\vb*{k})$ as function of $k_z$ for various values of
    $k_x,k_y$. The other parameters are the same as
    Fig.\,\ref{fig:Veff}. }
  \label{fig:omg_AFM-mag_FM}
\end{figure*}

\begin{figure*}[t]
  \begin{center}
    \includegraphics[scale=0.7]{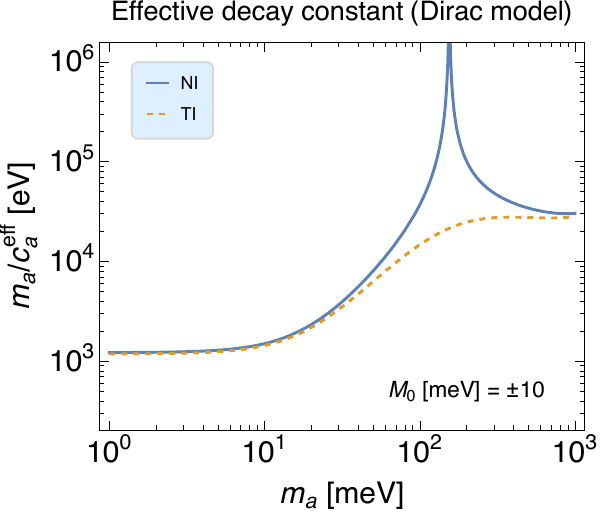}~~~~~
    \includegraphics[scale=0.7]{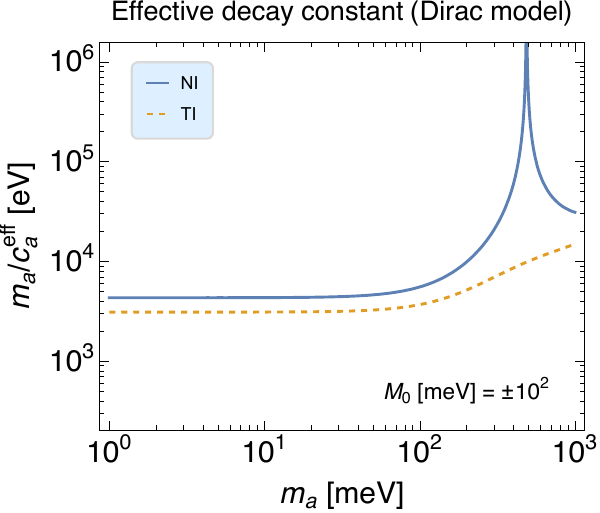}
  \end{center}
  \caption{Same as Fig.\,\ref{fig:feff-amp} computed in the Dirac
    model. $M_0\,[{\rm meV}]=\pm 10$ (left) and $\pm 10^2$ (right) are
    taken, where positive (negative) $M_0$ corresponds to the NI (TI)
    phases. The other parameters are taken as $A_1=A_2=1~{\rm eV}$. }
  \label{fig:feff-amp_Dirac}
\end{figure*}

\begin{figure*}[t]
  \begin{center}
    \includegraphics[scale=0.5]{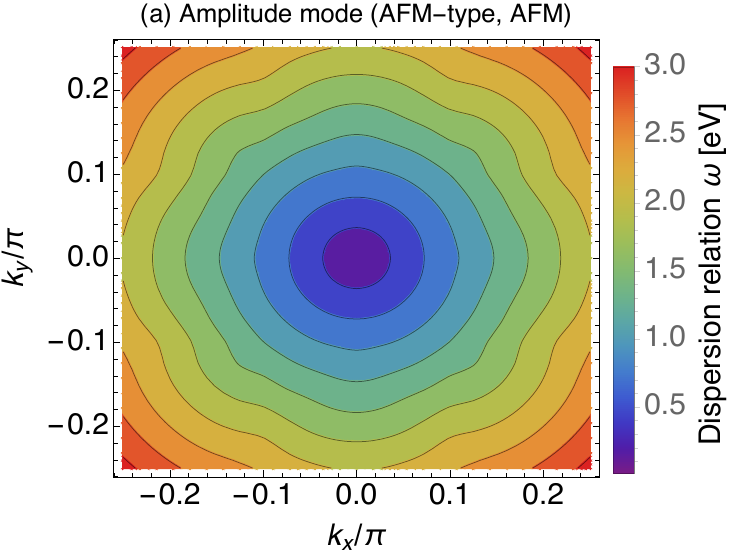}~~~~~
    \includegraphics[scale=0.5]{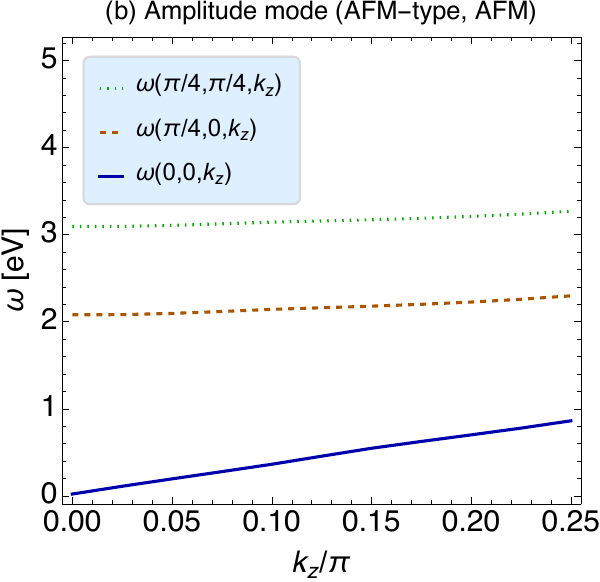}
  \end{center}
  \caption{Same as Fig.\,\ref{fig:omg_AFM-amp_AFM} but
    using another values for $A_1,A_2,B_1,B_2,M_0$ that are given
  by Ref.\,\cite{Li:2020fvr}.}
  \label{fig:omg_AFM-amp_AFM_Li}
\end{figure*}

\section{Additional figures}
\label{app:mag_FM}

The dispersion relations for $\alpha$- and $\beta$-type magnons under
the FM order are shown in Figs.\,\ref{fig:omg_FM-mag_FM} and
\ref{fig:omg_AFM-mag_FM}, respectively.

As the AFM order, we expect a gapless mode under the FM in the case of
$d_1=d_2=0$. In fact, we found the $\alpha$-type magnon becomes
gapless at $\vb*{k}=\vb*{0}$.  While it is gapless, the mode turns out
to be unstable.  Here `unstable' means
\begin{align}
  \tilde{\Gamma}^\alpha_F(0,\vb*{k})<0\,,
\end{align}
around $\vb*{k}=\vb*{0}$. Namely, the spatial kinetic term has the
opposite sign. Thus, such a mode is regarded as
unphysical.

Fig.\,\ref{fig:feff-amp_Dirac} shows the effective decay constant
computed in the Dirac model. The Dirac model is given by expanding
$d_a$ in Eq.\,\eqref{eq:d_a} around $\vb*{k}=0$.  Using the same
coefficients in the 3D TI model, we parametrize the Dirac model as
\begin{align}
  (d_1,d_2,d_3,d_4) =
  (A_2  k_x,\, A_2 k_y,
  \,A_1  k_z,\,M_0)\,.
  \label{eq:d_a_Dirac}
\end{align}

Finally, the dispersion relation of the AFM-type amplitude mode by
using the parameters given in Ref.\,\cite{Li:2020fvr} is given in
Fig.\,\ref{fig:omg_AFM-amp_AFM_Li}. As in
Fig.\,\ref{fig:omg_AFM-amp_AFM}, the stability of this excitation is
confirmed. In regions where $|k_x|$ and $|k_y|$ are large, the
contours behave differently, but this is due to the relatively large
value of $|B_2|$, which is affected by the boundary in the wavenumber
space integration.  The behavior around $\vb*{k}\sim 0$ is close to
linear dispersion, which is simply because the gap is much smaller
than eV.

\bibliography{draft}
\bibliographystyle{h-physrev5}

\end{document}